\documentclass[11pt]{article}

\usepackage{arxiv}
\usepackage[utf8]{inputenc}
\usepackage[T1]{fontenc}
\usepackage{lmodern}
\usepackage[colorlinks=true, linkcolor=snsblue, citecolor=snsgreen, urlcolor=snsred]{hyperref}
\usepackage{url}
\usepackage{booktabs}
\usepackage{amsmath}
\usepackage{amssymb}
\usepackage{amsfonts}
\usepackage{nicefrac}
\usepackage{microtype}
\usepackage{natbib}
\usepackage{siunitx}
\usepackage{mathtools}
\usepackage{algorithm}
\usepackage{algpseudocode}
\usepackage{csquotes}
\usepackage[nameinlink]{cleveref}
\usepackage{ulem}

\usepackage{xcolor}
\definecolor{snsblue}{RGB}{76, 114, 176}
\definecolor{snsgreen}{RGB}{85, 168, 104}
\definecolor{snsred}{RGB}{196, 78, 82}
\definecolor{snspurple}{RGB}{129, 114, 178}
\definecolor{snsyellow}{RGB}{204, 185, 116}
\definecolor{snscyan}{RGB}{100, 181, 205}

\newcommand{\ie}{i.e.}

\title{Impact of an Ensemble of Ocean Data Assimilations in ECMWF's next generation ocean reanalysis system}

\author{
Marcin Chrust\\
ECMWF\\
Shinfield Park\\
Reading, United Kingdom\\
\texttt{marcin.chrust@ecmwf.int}\\
\And
Anthony T. Weaver\\
CERFACS / CECI CNRS UMR 5318\\
Toulouse, France\\
\AND
Philip Browne\\
ECMWF\\
Shinfield Park\\
Reading, United Kingdom\\
\And
Hao Zuo\\
ECMWF\\
Shinfield Park\\
Reading, United Kingdom\\
\And
Magdalena Alonso Balmaseda\\
ECMWF\\
Shinfield Park\\
Reading, United Kingdom\\
}

\begin{document}

\maketitle

\begin{abstract}

An Ensemble of Data Assimilations (EDA) can provide valuable information on the analysis and short-range forecast uncertainties. The present ECMWF operational ocean analysis and reanalysis system, called ORAS5, produces an ensemble but does not exploit it for the specification of the background-error covariance matrix $\mathbf{B}$, a key component of the data assimilation system. In this article, we describe EDA developments for the ocean, which take advantage of the short-range forecast ensemble for specifying, in two distinct ways, parameters of a  covariance model representation of $\mathbf{B}$. First, we generate a climatological ensemble over an extended period to produce seasonally varying climatological estimates of background-error variances and horizontal correlation length-scales. Second, on each assimilation cycle, we diagnose flow-dependent variances from the ensemble and blend them with the climatological estimates to form hybrid variances. We also use the ensemble to  diagnose flow-dependent vertical correlation length-scales. We demonstrate for the Argo-rich period that this new, hybrid formulation of $\mathbf{B}$ results in a significant reduction of background errors compared to the parameterized formulation of $\mathbf{B}$ used in ORAS5. The new ocean EDA system will be employed in ORAS6, ECMWF's next generation ocean  reanalysis system.

\end{abstract}

\keywords{hybrid data assimilation, ensemble of data assimilations, variational assimilation, background error covariances}

\section{Introduction}

The current generation of the ECMWF ocean reanalysis system, called ORAS5 \citep{zuo-2019}, is an ensemble system, but makes no use of the ensemble of forecasts to provide climatological or flow-dependent covariance information on forecast errors. Such information could be used to improve the specification of the background-error covariance matrix $\mathbf{B}$ in the variational data assimilation system used to produce the ocean reanalysis. This is supported by growing evidence of the benefits of using ensembles to enrich the representation of $\mathbf{B}$ not only in variational data assimilation systems for the atmosphere, where it is already a very well-established practice \citep{buehner-2005, bonavita-2012, clayton-2013, berre-2015, kleist-2015, bonavita-2016}, but also for the ocean \citep{daget-2009, penny-2015, storto-2018, lea-2022}. 

In this article, we describe an Ensemble of Data Assimilations (EDA) framework that will be the basis of the upcoming 6th generation of ECMWF's ocean reanalysis system (ORAS6). While the ensemble of forecasts can be used directly to define a sample covariance matrix estimate of $\mathbf{B}$ (e.g., see \cite{lorenc-2003} and \cite{buehner-2005}), here we use it indirectly to calibrate the parameters of a modelled representation of $\mathbf{B}$. To simplify the illustration of how this is achieved, we follow the standard decomposition of a covariance matrix into a correlation matrix and a diagonal matrix of standard deviations that multiplies the correlation matrix on either side. The ensemble of forecasts are used in two distinct ways. First, using the empirically parameterized representation of $\mathbf{B}$ from ORAS5, an $11$-member ensemble is produced for the years 2010--2015 in order to compute seasonal climatological estimates of both standard deviations and parameters of the correlation matrix. Second, the ensemble of forecasts is also employed to estimate flow-dependent forecast errors, by updating on each assimilation cycle the background-error standard deviations and vertical correlation length-scales. We then conduct a series of data assimilation experiments for the period 2017--2021 and assess the strengths and weaknesses of the new formulation of $\mathbf{B}$. We demonstrate that accounting for flow-dependent background-error covariances allows surface observations to be assimilated more effectively, including, for the first time at ECMWF, sea surface temperature (SST) observations. We show that the improvements in $\mathbf{B}$ coupled with the assimilation of SST have enabled us to address a long-standing issue of large temperature biases in the Gulf Stream that were present in ORAS5. 

The development of the ocean EDA can also be seen as an important step towards building a coupled EDA system at ECMWF. It is expected \citep{rosnay-2022} that coupled data assimilation would allow for better exploitation of interface observations and for reducing the initialization imbalances of the Earth System Model. The ability of the ocean data assimilation system to exploit surface observations in an optimal way is thus of strategic importance.

The article is structured as follows. \Cref{sec:methodology} describes methodological aspects of the new system. An overview of the covariance model used to define $\mathbf{B}$ is presented in \Cref{subsec:b} and is followed in \Cref{subsec:bens} by a description of how the forecast ensemble is used to estimate the covariance model parameters. \Cref{subsec:bhybclim} describes how the climatologies of the covariance model parameters are constructed and subsequently combined with ``errors of the day'' to form a hybrid covariance model. The performance of the new system is discussed in \Cref{sec:results}. First, a baseline configuration, defined using an ORAS5-like $\mathbf{B}$, is described in \Cref{subsec:boras5}. The subsequent subsections illustrate the impact of using ensemble-derived parameters in $\mathbf{B}$. \Cref{subsec:bhybvar-res} and \Cref{subsec:bhybten-res} describe the impact of the hybrid variances and hybrid correlation model, respectively, while \Cref{subsec:bhybtenvar-res} describes the impact of the fully hybrid covariance model that includes both the hybrid variances and hybrid correlation model.
Conclusions and outlook are given in \Cref{sec:conclusions}.

\section{Methodology} \label{sec:methodology}

The new ocean EDA system is built around NEMO v4.0.6, coupled to a sea-ice model SI$^{3}$ \citep{madec-2023}. The model employs a tri-polar extended ORCA grid at approximately $0.25^{\circ}$ horizontal resolution and with $75$ vertical levels. The EDA consists of an unperturbed control member accompanied by $10$ perturbed members. The number of perturbed members was chosen based on the consideration of computational cost and the requirement to reduce sampling noise when estimating the parameters of $\mathbf{B}$.  The perturbed members assimilate randomly displaced observations and are forced using hourly perturbed forcing fields derived from ERA5 \citep{hersbach-2020}. The ocean initial conditions are implicitly perturbed via the cycling process of the EDA. The model itself remains unperturbed as in ORAS5. Data assimilation is carried out using NEMOVAR \citep{mogensen-2012} in its three-dimensional variational assimilation (3D-Var) First Guess at Appropriate Time (FGAT) configuration with a $5$-day assimilation window. The analysis increment obtained from the 3D-Var minimization is applied to the model as a tendency from the start of the window using Incremental Analysis Updates \citep{bloom-1996}. The assimilated observations contain information on temperature, salinity, sea surface height (SSH) and sea ice concentration (SIC). The details of the observing network and perturbation schemes are outside the scope of this article. The interested reader is referred to the upcoming ORAS6 article for more details \citep{zuo-2024}. Here, we focus on the description of $\mathbf{B}$  and the methods for using the ensemble of forecasts to compute climatological and flow-dependent parameters for $\mathbf{B}$. Details of the $\mathbf{B}$ formulation employed in the current operational ocean reanalysis system, ORAS5, are provided to facilitate comparison and interpretation of the results. The impact of the new $\mathbf{B}$ formulation is illustrated in the data-rich period 2017--2021.

    \subsection{The background-error covariance model} \label{subsec:b}

In this section, we describe the formulation of $\mathbf{B}$ that is used for ORAS6 and in doing so expose the covariance model parameters that are estimated using
the EDA. We keep technical details to a minimum as many of them can be found in
existing references \citep{weaver-2005, balmaseda-2013, weaver-2016, weaver-2021}.

The ocean model state variables consist of potential temperature ($T$), practical salinity ($S$),
the two components of horizontal velocity ($u$ and $v$), SSH ($\eta$) and SIC ($a$).
In the ORAS6 3D-Var FGAT system, the control variables that are estimated through
data assimilation are temperature, SIC, and the so-called {\it unbalanced} parts of
salinity ($S_{\rm u}$) and SSH ($\eta_{\rm u}$). The unbalanced parts of
horizontal velocity ($u_{\rm u}$ and $v_{\rm u}$) are not considered as
control variables since no information is assimilated in the
3D-Var system to correct them.
The state and control vectors of the assimilation problem consist of
discrete values of the state and control variables at the model grid-points.

\subsubsection{Balance operator}

The background-error covariance matrix is formulated in terms of a
balance operator $\mathbf{K}_{\rm b}$ and a covariance matrix $\mathbf{B}_{\rm c}$
for the control variables:
\begin{equation}
  \mathbf{B} \, = \, \mathbf{K}_{\rm b} \, \mathbf{B}_{\rm c} \, \mathbf{K}_{\rm b}^{\rm T}.
  \nonumber
\end{equation}
We make the fundamental assumption that the
control variables are mutually uncorrelated, so that $\mathbf{B}_{\rm c}$
can be represented by a block-diagonal matrix
\begin{equation}
  \mathbf{B}_{\rm c} =
  \! \left( \! \!
\begin{array}{cccc}
  \mathbf{B}_{TT} & \mathbf{0}       & \mathbf{0}   & \mathbf{0} \\
  \mathbf{0}      & \mathbf{B}_{S_{\rm u}S_{\rm u}}  &  \mathbf{0}  & \mathbf{0} \\
  \mathbf{0}      & \mathbf{0}       & \mathbf{B}_{\eta_{\rm u} \eta_{\rm u}} & \mathbf{0} \\
  \mathbf{0}      & \mathbf{0}       & \mathbf{0}   & \mathbf{B}_{aa}
\end{array}
\! \! \right),
  \nonumber
\end{equation}
where $\mathbf{B}_{\alpha \alpha}$ is a univariate covariance matrix
for the control variable \mbox{$\alpha = T$},  $S_{\rm u}$, $\eta_{\rm u}$ or $a$.

Cross-covariances between the background errors of different state variables
are induced by the balance operator, which has the specific structure \citep{weaver-2005}
\begin{equation}
  \mathbf{K}_{\rm b} =
  \! \left( \! \!
\begin{array}{cccc}
  \mathbf{I}_{N_{\rm hz}} &  \mathbf{0}             & \mathbf{0}               & \mathbf{0}  \\
  \mathbf{K}_{ST}       &  \mathbf{I}_{N_{\rm hz}}   & \mathbf{0}              & \mathbf{0}  \\
  \mathbf{K}_{\eta T}   & \mathbf{0}  & \mathbf{I}_{N_{\rm h}}     & \mathbf{0}  \\
  \mathbf{0}           & \mathbf{0}               & \mathbf{0}               & \mathbf{I}_{N_{\rm h}} \\
  \mathbf{K}_{u T}      & \mathbf{0}   & \mathbf{K}_{u \eta_{\rm u}}  & \mathbf{0}  \\
  \mathbf{K}_{v T}      & \mathbf{0}   & \mathbf{K}_{v \eta_{\rm u}}  & \mathbf{0}
\end{array}
\! \! \right)
\label{eq:Kb}
\end{equation}
where the block components $\mathbf{K}_{\beta \alpha}$ constitute the balance
between the control variable $\alpha$ and the state variable $\beta$, and
$\mathbf{I}_{N_{\rm h}}$ and $\mathbf{I}_{N_{\rm hz}}$ are identity matrices,
$N_{\rm h}$ and $N_{\rm hz}$ corresponding to the
number of two-dimensional (2D; horizontal) and
three-dimensional (3D; horizontal-vertical) grid-points, respectively.
The specific balance relationships ($\mathbf{K}_{\beta \alpha}$)
between variables are essentially those described in \cite{ricci-2005}, \cite{weaver-2005} and \cite{balmaseda-2013}, but
with some refinements that have been introduced over the years. The basic relations are summarized below.

The balance between $T$ and $S$ ($\mathbf{K}_{ST}$) involves making a local vertical adjustment to $S$ to approximately preserve the water-mass properties of the background state \citep{ricci-2005}. Changes to the water-mass properties
occur through the unbalanced component $S_{\rm u}$. The other variables ($\eta$, $u$ and $v$) depend on $T$ and $S$ through density, which is defined from a linear equation of state with thermal expansion and saline contraction coefficients computed from the background state. Note that only the balanced component of $S$ is used in the computation of density, which is reflected by the absence of balance between $S_{\rm u}$ and the other variables in Equation~(\ref{eq:Kb}). The balance between $\eta$ and $T$ ($\mathbf{K}_{\eta T}$) is determined by computing dynamic height at the surface relative to the ocean bottom. The unbalanced component $\eta_{\rm u}$ is interpreted as a barotropic contribution to SSH, that is in balance with a depth-independent geostrophic velocity component ($\mathbf{K}_{u \eta_{\rm u}}$ and $\mathbf{K}_{v \eta_{\rm u}}$). The remaining velocity balance ($\mathbf{K}_{uT}$ and $\mathbf{K}_{vT}$) is depth-dependent and assumed to be in geostrophic balance with a horizontal gradient of pressure computed from the hydrostatic relation. Near the Equator, the $f$-plane geostrophic balance  is smoothly reduced to zero, and replaced by a $\beta$-plane geostrophic balance for the $u$-component only.

In Equation~(\ref{eq:Kb}) the velocity balance acts as a strong constraint since there are no unbalanced velocity components
in the control vector.
It would be sufficient then to apply this particular balance directly
to the analysis increments of $T$, $S_{\rm u}$ and
$\eta_{\rm u}$ generated by the 3D-Var minimization.
However, instead of applying the velocity balance as a post-analysis adjustment,
we have formally included it in $\mathbf{K}_{\rm b}$
(and hence as part of the 3D-Var minimization algorithm) in preparation for future extensions of the system that will
assimilate velocity information and thus allow us to estimate the unbalanced (ageostrophic) components of velocity.
There is currently no balance imposed between SIC and any of the other variables.

\subsubsection{The univariate covariance model}

The block covariance matrices in $\mathbf{B}_{\rm c}$ are factored as
\begin{equation}
  \mathbf{B}_{\alpha \alpha} \, = \, \boldsymbol{\Sigma}_{\alpha} \,
  \mathbf{C}_{\alpha \alpha}\, \boldsymbol{\Sigma}_{\alpha}
  \nonumber
\end{equation}
where $\mathbf{C}_{\alpha \alpha}$ is a correlation matrix and $\boldsymbol{\Sigma}_{\alpha}$ is
a diagonal matrix of standard deviations. 
The standard deviations for all control variables except for SIC
are estimated using the EDA as described in the next section.
Each correlation matrix is modelled using
a diffusion operator, formed by discretising the pseudo-time derivative of a
diffusion equation with an implicit (Euler backward) scheme.
In what follows, we refer to this as an implicit diffusion operator.
The total number of diffusion steps $M$ primarily controls the spectral decay rate of the smoothing kernel of the diffusion operator at high wavenumbers.
These smoothing
kernels are covariance functions from the Mat\'ern family \citep{guttorp-2006,weaver-2013}.
Here, we have set $M$ to a value of 10 (for all variables) so that
the spatial correlations are approximately Gaussian (see Figure~A1 in \cite{goux-2024}).

The correlation matrix has the symmetric, factored form
\begin{equation}
  \mathbf{C}_{\alpha \alpha} \, = \, \boldsymbol{\Gamma}_{\alpha } \, \mathbf{V}_{\alpha}\,
  \mathbf{W}_{\alpha}^{-1}\, \mathbf{V}_{\alpha}^{\rm T} \,
  \boldsymbol{\Gamma}_{\alpha}
  \label{eq:C}
\end{equation}
where $\mathbf{V}_{\alpha}$ is a discrete representation of a diffusion operator
as detailed below;
$\boldsymbol{\Gamma}_{\alpha}$ is a diagonal matrix containing normalization factors
to ensure that the diagonal elements of $\mathbf{C}_{\alpha \alpha}$ are all approximately
equal to one; and
$\mathbf{W}_{\alpha}$ is a diagonal matrix of either area elements ($\mathbf{W}_{\rm h}$)
or volume elements ($\mathbf{W}_{\rm hz}$) depending on the variable:
\begin{equation}
  \mathbf{W}_{\alpha} =
  \left\{
  \begin{array}{lcl}
    \mathbf{W}_{\rm h} & \mbox{if}  & \alpha = \eta_{\rm u} \; \mbox{or} \; a,
   \\
     \mathbf{W}_{\rm hz} & \mbox{if}  & \alpha = T \; \mbox{or} \; S_{\rm u}.
  \end{array}
  \right.
  \nonumber
\end{equation}

The combined matrix operator $\mathbf{V}_{\alpha} \mathbf{W}_{\alpha}^{-1} \mathbf{V}_{\alpha}^{\rm T}$
represents the full $M$-step diffusion process, while the components $\mathbf{V}_{\alpha}$
and $\mathbf{V}_{\alpha}^{\rm T}$ constitute diffusion over half the number of steps.
Horizontal and vertical correlations are
modelled using a 2D and one-dimensional (1D) implicit diffusion
operator, respectively.
Horizontal-vertical (3D) correlations are modelled by combining the horizontal
and vertical implicit diffusion operators. Within $\mathbf{V}_{\alpha}$, the combined
operator, which is used for $T$ and $S_{\rm u}$,
is formed by first applying one step of the 1D implicit diffusion operator
at each horizontal grid-point (an operation denoted by $\mathbf{F}_{\alpha, {\rm z}}$),
followed by applying one step of the 2D implicit diffusion operator in each vertical
level (an operation denoted by $\mathbf{F}_{\alpha, {\rm h}}$).  We refer to this 3D diffusion
formulation as 2D$\times$1D. For $\eta_{\rm u}$ and $a$,
only the horizontal diffusion operator is needed. In terms of the one-step components, $\mathbf{V}_{\alpha}$ can be expressed as
\begin{equation}
  \mathbf{V}_{\alpha} =
  \left\{
  \begin{array}{lcl}
    \mathbf{F}_{\alpha, {\rm h}}^{M/2} & \mbox{if} & \alpha = \eta_{\rm u} \; \mbox{or} \; a,
   \\
   \left( \mathbf{F}_{\alpha, {\rm h}} \, \mathbf{F}_{\alpha, {\rm z}}
   \right)^{M/2} & \mbox{if} &  \alpha = T \; \mbox{or} \; S_{\rm u}.
  \end{array}
  \right.
  \label{eq:V_alpha}
\end{equation}

Rather than constructing the 3D correlation operator as a product
of 2D and 1D diffusion operators as in Equation~(\ref{eq:V_alpha}), an alternative
and arguably more natural approach is to construct it directly from a discretised
3D implicit diffusion equation (see Section~2.2.3 in \cite{weaver-2021}). The resulting correlation
model has similar smoothness properties to Equation~(\ref{eq:V_alpha})
(see Figure~18 in \cite{weaver-2021}) but is computationally more expensive
and therefore has not been used for ORAS6.

Consider the application of the 2D$\times$1D implicit diffusion operator to a vector
$\boldsymbol{\alpha}_{m-1}$ where the subscript $m$ denotes
the step counter in the diffusion process.
Performing one step of 2D implicit diffusion involves solving a linear system
\begin{equation}
  \mathbf{A}_{{\rm h}_k} \boldsymbol{\alpha}_m^{{\rm h}_k} = \boldsymbol{\alpha}_{m-1}^{{\rm h}_k}
  \label{eq:2D_linsys}
\end{equation}
where $\boldsymbol{\alpha}_{m-1}^{{\rm h}_k}$ is the sub-vector of $\boldsymbol{\alpha}_{m-1}$
associated with model level $k$.
Similarly, performing one step of 1D implicit diffusion involves solving,
at each horizontal grid-point $(i,j)$, a linear system
\begin{equation}
\mathbf{A}_{{\rm z}_{ij}} \boldsymbol{\alpha}_m^{{\rm z}_{ij}} = \boldsymbol{\alpha}_{m-1}^{{\rm z}_{ij}}
  \label{eq:1D_linsys}
\end{equation}
where $\boldsymbol{\alpha}_{m-1}^{{\rm z}_{ij}}$ is the sub-vector of $\boldsymbol{\alpha}_{m-1}$
associated with the vertical column at $(i,j)$.
The matrices $\mathbf{A}_{{\rm h}_k}$ and $\mathbf{A}_{{\rm z}_{ij}}$ are positive definite
and self-adjoint with respect to the inner product
$\boldsymbol{\alpha}_i^{\rm T} \mathbf{W}_{\rm hz} \boldsymbol{\alpha}_j$.

The matrices $\mathbf{A}_{{\rm h}_k}$ and $\mathbf{A}_{{\rm z}_{ij}}$ are constructed
from the horizontal and vertical components of the 3D elliptic operator
associated with the implicitly discretised 3D diffusion equation. 
In NEMOVAR, the analysis is performed directly on the ORCA grid, and the
spatial discretisation techniques used for the diffusion equation are inherited from NEMO  \citep{madec-2023}.
Adopting NEMO notation, the continuous counterparts of
$\mathbf{A}_{{\rm h}_k}$ and $\mathbf{A}_{{\rm z}_{ij}}$ are the elliptic operators
\begin{equation}
  A_{{\rm h}} :=
  I - \frac{1}{e_1 e_2 e_3}
   \left( \frac{\partial}{\partial i}\left( \kappa_{11} \frac{e_2 e_3}{e_1} \frac{\partial}{\partial i} \right)
 +        \frac{\partial}{\partial j}\left( \kappa_{22} \frac{e_1 e_3}{e_2} \frac{\partial}{\partial j} \right) \right)
  \label{eq:2D_diffusion}
\end{equation}
and
\begin{equation}
  A_{{\rm z}} :=
  I - \frac{1}{e_1 e_2 e_3}
   \left( \frac{\partial}{\partial k}\left( \kappa_{33} \frac{e_1 e_2}{e_3} \frac{\partial}{\partial k} \right) \right)
  \label{eq:1D_diffusion}
\end{equation}
where $I$ is the identity operator, and $e_1$, $e_2$ and $e_3$ are the scale factors that define the
local deformation of the general orthogonal curvilinear coordinates $(i,j,k)$ in the $i$, $j$ and $k$ direction, respectively. Note that the factor $e_1 e_2 e_3$ corresponds to a volume element and its value at each grid-point constitutes
a diagonal element of $\mathbf{W}_{\rm hz}$. The elliptic operator
associated with the 2D diffusion-based correlation operators for $\eta_{\rm u}$ and $a$ is the same
as Equation~(\ref{eq:2D_diffusion}) but with the scale factor $e_3$ omitted.
The diffusion coefficients $\kappa_{11}$ and $\kappa_{22}$
in Equation~(\ref{eq:2D_diffusion}) are the parameters that control the length-scales
of the horizontal correlations in the $i$ and $j$ directions,
while the diffusion coefficient $\kappa_{33}$ is the parameter that controls the length-scale
of the vertical correlations. These parameters depend on the three spatial coordinates $(i,j,k)$.
More generally, these coefficients constitute the diagonal elements of a 3D diffusion tensor at
each grid-point.
In the next section, we describe how these parameters are estimated from the EDA.

Equation~(\ref{eq:1D_linsys}) is solved exactly at each grid-point $(i,j)$ by decomposing $\mathbf{A}_{{\rm z}_{ij}}$ into Cholesky factors and using forward and backward substitution. 
Solving Equation~(\ref{eq:2D_linsys}) is the most computationally demanding operation
in $\mathbf{B}$. In NEMOVAR, it is solved approximately in each level $k$ using a linear solver
based on the Chebyshev Iteration \citep{weaver-2016}. The same, fixed number of iterations are
used in $\mathbf{V}_{\alpha}$ and $\mathbf{V}_{\alpha}^{\rm T}$ in order to preserve the
symmetric property of $\mathbf{C}_{\alpha \alpha}$
to within machine precision. The number of iterations for each diffusion step
is pre-computed by solving the 2D implicit diffusion equation with a random right-hand side
and diagnosing the iterations required to achieve a three orders of magnitude reduction
of the 2-norm of the residual relative to the 2-norm of the right-hand side.
The total number of iterations ranges between 8 and 24 for $T$ and $S_{\rm u}$, and 25 and 36 for $\eta_{\rm u}$ and $a$.

    \subsection{Estimation of variances and correlation model parameters} 
    \label{subsec:bens}

  The ensemble perturbations from the EDA are used to calibrate parameters of the background-error covariance models for $T$, $S_{\rm u}$ and $\eta_{\rm u}$.
  These parameters consist of the standard deviations $\sigma$ and the coefficients $\kappa_{11}$, $\kappa_{22}$ and $\kappa_{33}$
  of the diffusion-based correlation model, at each grid-point. 
  The total number of discrete parameters to estimate is
  2~variables $\times$ 4~parameters $\times$ $N_{\rm hz}$ grid-points $+$ 1~variable $\times$ 3~parameters $\times$ $N_{\rm h}$
  grid-points,
  where $N_{\rm hz}$ and $N_{\rm h}$ are understood here to include only the active (ocean) points. In the ORAS6 configuration,
  this amounts to approximately \mbox{$4.3\times 10^8$} discrete parameters.

  The inverse of the balance operator is required to compute perturbations of the control variables from the perturbations of the state variables provided by the ensemble. In particular, we need to compute 
  the perturbations of $T$, $S_{\rm u}$ and $\eta_{\rm u}$
  by removing the balanced component from the perturbations of $T$, $S$ and $\eta$. This computation requires applying the inverse of the upper 3$\times$3 block matrix in Equation~(\ref{eq:Kb}), which is straightforward given its lower triangular structure. Let $\widehat{\mathbf{K}}_{\rm b}$ denote this block sub-matrix and let $\left\{ \mathbf{x}_p  \right\}$, $p=0,\ldots, N_{\rm e}$, denote the ensemble of background
  states of $T$, $S$ and $\eta$ where $p=0$ corresponds to the unperturbed member.
  If $\mathbf{e}_p^{\prime}$ is the vector containing the perturbations of $T$, $S$ and $\eta$, centred about their mean,
  \begin{equation}
  \mathbf{e}_p^{\prime} = \mathbf{x}_p - \frac{1}{N_{\rm e}}\sum_{l=1}^{N_{\rm e}} \mathbf{x}_l,
  \nonumber
  \end{equation}
then the centred control vector perturbations $\boldsymbol{\epsilon}_p^{\prime}$ are defined as
    \begin{equation}
    \boldsymbol{\epsilon}_p^{\prime} = \widehat{\mathbf{K}}_{\rm b}^{-1} \mathbf{e}_p^{\prime},
     \hspace{1cm}     p=1,\ldots, N_{\rm e}.
    \label{eq:pert}
  \nonumber
  \end{equation}
  The unperturbed member $\mathbf{x}_0$ is used as the reference state for the \mbox{$T$-$S$} and density relationships in $\widehat{\mathbf{K}}_{\rm b}$ and therefore is excluded from the ensemble average.

  A sample estimate of the variances of $T$, $S_{\rm u}$ and $\eta_{\rm u}$ is computed
  on each cycle from the \mbox{$N_{\rm e} = 10$} ensemble perturbations $\boldsymbol{\epsilon}_p^{\prime}$. To reduce sampling error, the 
  variances are filtered in each level using a 2D implicit diffusion operator with a horizontal filtering
  length-scale determined using the method based on optimality criteria for linear filtering of covariances derived by \cite{menetrier-2015}. The procedure
  for computing a sample estimate of the diffusion coefficients is more complicated and is detailed
  in Section~3 of \cite{weaver-2021}. Only the main points are summarized here.

Central to the method is the relationship between Mat\'ern correlation functions and the smoothing kernels of the implicit diffusion operator.
The local curvature of the correlation function near its peak can be used to characterize the spatial range of the function. For an anisotropic function, the local curvature is defined by the Local Correlation (Hessian) Tensor (LCT). If $\boldsymbol{H}$ denotes the LCT of an at least twice differentiable correlation function $c(r)$ then $\boldsymbol{H}:= -\nabla \nabla^{\rm T} c |_{r=0}$ where $r$ is Euclidean distance. For Mat\'ern functions in $\mathbb{R}^2$, we have the relation 
\citep{weaver-2013}
\begin{equation}
  \boldsymbol{\kappa} \, = \, \left( \frac{1}{2M-4} \right)\boldsymbol{H}^{-1}
  \label{eq:k_2d}
\end{equation}
where $\boldsymbol{\kappa}$ is the diffusion tensor of a 2D implicit diffusion operator. The LCT inverse, $\boldsymbol{H}^{-1}$, for anisotropic functions can be interpreted as the tensor generalisation of the Daley length-scale for isotropic correlation functions \citep{daley-91}.

In 2D, $\boldsymbol{H}$ is a \mbox{$2\times 2$} symmetric, positive-definite matrix with elements
$H_{pq}$, \mbox{$p,q = 1, 2$}. This matrix can be approximated at each grid point from a sample estimate of the tensor product of the local gradient of the ensemble perturbations normalized by their sample standard deviation \citep{michel-2016}:
\begin{equation}
  \boldsymbol{H}
    \, \approx \, \widetilde{\boldsymbol{H}} = \frac{1}{N_{\rm e} - 1} 
    \sum_{p=1}^{N_{\rm e}} \nabla \big( \epsilon^{\prime}_p / \widetilde{\sigma} \big)\,
      \big( \nabla \big( \epsilon^{\prime}_p/ \widetilde{\sigma} \big)\, \big)^{\rm T}
  \label{eq:H_ens}
\end{equation}
where
\begin{equation}
  \widetilde{\sigma}
    \, = \, \sqrt{\frac{1}{N_{\rm e} - 1} 
    \sum_{p=1}^{N_{\rm e}} {\epsilon^{\prime}_p}^{2}}.
  \label{eq:sigma}
  \nonumber
\end{equation}
The derivatives in Equation~(\ref{eq:H_ens}) are estimated numerically using centred finite-differences as described in \cite{weaver-2021}. As with the variances, each of the estimated LCT elements are filtered in each model level using a 2D diffusion operator with a constant horizontal filtering length-scale, where the procedure of \cite{michel-2016} has been adopted in order to preserve positive definiteness of the LCT. After filtering, the LCT is inverted at each grid-point to obtain the Daley tensor and hence an estimate of the diffusion tensor from Equation~(\ref{eq:k_2d}). As evident from Equation~(\ref{eq:2D_diffusion}), the current 2D implicit diffusion operator does not account for the cross-derivative terms that are needed to exploit the non-diagonal elements of the tensor. To compensate, the diagonal elements are re-scaled such that the determinant of the adjusted diagonal tensor equals the determinant of the estimated non-diagonal tensor:
\begin{eqnarray}
\left.
\begin{array}{ccl}
  \kappa_{11} & \! \! = \! \! & \widetilde{\kappa}_{11} \,
  \sqrt{ 1 - \widetilde{\kappa}_{12}^2 / \widetilde{\kappa}_{11} \widetilde{\kappa}_{22}}
\vspace{3mm}
\\
\kappa_{22} & \! \! = \! \! & \widetilde{\kappa}_{22} \,
\sqrt{ 1 - \widetilde{\kappa}_{12}^2 / \widetilde{\kappa}_{11} \widetilde{\kappa}_{22}}
\end{array}
\right\}
\nonumber
\end{eqnarray}
where the diffusion coefficients with the tilde are the original estimates. The final values of $\kappa_{11}$ and $\kappa_{22}$ are obtained after applying procedures to bound their minimum-allowed values and to reduce their values in semi-enclosed seas and close to boundaries where the filtered sample estimates could be excessively large \citep{weaver-2021}.

A similar estimation and filtering procedure is used to determine the vertical diffusion coefficients for the 1D diffusion operator (Equation~(\ref{eq:1D_diffusion})). Instead of Equation~(\ref{eq:k_2d}), we make use of the relation
\begin{equation}
  \kappa_{33} \, = \, \left( \frac{1}{2M-3} \right) {H}_{33}^{-1},
  \label{eq:k33}
  \nonumber
\end{equation}
which has a slightly modified scaling factor that is appropriate for Mat\'ern functions in $\mathbb{R}$. The element $H_{33}^{-1}$ can be interpreted as the square of the vertical (Daley) length-scale. It is estimated at each grid-point from the inverse of a sample estimate of the square of the vertical derivative of the ensemble perturbations normalized by their sample standard deviation:
\begin{equation}
  H_{33}
    \, \approx \, \widetilde{H}_{33} = \frac{1}{N_{\rm e} - 1} 
    \sum_{p=1}^{N_{\rm e}} \left(\frac{1}{e_{3}} \frac{\partial
    ( \epsilon^{\prime}_p/ \widetilde{\sigma} ) }{\partial k} \right)^{\! 2}.
  \label{eq:H33_ens}
  \nonumber
\end{equation}
Before inverting $\widetilde{H}_{33}$, the estimate is filtered using a 1D diffusion operator equipped with an optimally-estimated vertical filtering length-scale \citep{menetrier-2015}. Finally,  $\kappa_{33}$ is adjusted to bound the minimum-allowed value, and to avoid excessively large values in the mixed layer and in shallow-sea regions, as explained in \cite{weaver-2021}.

    \subsection{Hybrid $\mathbf{B}$ parameters} \label{subsec:bhybclim}

    In order to produce more robust estimates of the covariance model parameters, we have developed a procedure to hybridize the flow-dependent ensemble estimates with climatological and parameterized values. Hybrid variances are built by combining climatological, flow-dependent and parameterized variances, while the hybridization of the diffusion coefficients is defined somewhat differently. Here, it involves using climatological horizontal diffusion coefficients in the 2D diffusion operator and flow-dependent vertical diffusion coefficients in the 1D diffusion operator. In what follows, we will loosely refer to this way of combining climatological and flow-dependent diffusion coefficients as the {\it hybrid diffusion tensor} formulation. Climatological statistics were computed over the 6--year period 2010--2015 for each season using an 11-member (10 perturbed + 1 unperturbed) ensemble system equipped with the previous generation, parameterized  $\mathbf{B}$ (\ie~from ORAS5). The statistics were obtained by aggregating the filtered variances and filtered LCT elements across all the assimilation cycles in a season across the considered years. On average, the total effective ensemble size to compute the seasonal climatology was 1080 (10~perturbed member $\times$ 18~cycles per season $\times$ 6~years). Finally, the climatological diffusion coefficients are adjusted using the same procedures described in \Cref{subsec:bens} for the cycle-dependent diffusion coefficients.

        \subsubsection{Hybrid variances} \label{subsec:bhybclim-var}

        The background-error variances for $T$ are parameterized in ORAS5 \citep{zuo-2015} 
        in terms of the vertical derivative of the background temperature field in order to concentrate the largest variances at the level of the thermocline. The background-error variances for $S_{\rm u}$ are parameterized using the background mixed-layer depth in order to focus large variances in the mixed layer where the T-S balance relationship is not applied. The background-error variances for $\eta_{\rm u}$ are parameterized using a latitude dependent function to account for the importance of the barotropic component in the extra-tropics. 
        
        As noted earlier, the employed ensemble generation scheme still relies on the deterministic model and the only sources of uncertainty that it reflects are those attributed to representativeness error of observations and the uncertainty in surface forcing. For that reason, and also owing to the eddy-permitting horizontal resolution of the model not being sufficient to resolve ocean variability at the eddy scales and below, relying on ensemble statistics alone for the specification of the background-error variances is not possible. The ensemble spread outside of eddy-active regions, like the Western Boundary Current (WBC) or the Antarctic Circumpolar Current (ACC) regions, is very small and not representative of diagnosed background forecast errors. To mitigate this issue, we combine the seasonal climatological standard deviations with the ORAS5 parameterized standard deviations using a smooth maximum function:
         \begin{equation}
            \label{eq:smoothmax_m}
            \sigma_{\textrm{m}} = \frac{1}{h} \log \left( e^{h w_{\textrm{c}} \sigma_{\textrm{c}}} + e^{h w_{\textrm{p}} \sigma_{\textrm{p}}} - 1 \right)
        \end{equation}
        where $w_\textrm{c}$ and $w_\textrm{p}$ are dimensionless weighting factors for the climatological and parameterized standard deviations, $\sigma_\textrm{c}$ and $\sigma_\textrm{p}$, respectively, and $h$ is a hardness factor,
        which has the property that as $h\rightarrow \infty$, $\sigma_{\textrm{m}}$ approaches the hard maximum, $\max \big( w_{\textrm{c}} \sigma_{\textrm{c}}, w_{\textrm{p}} \sigma_{\textrm{p}} \big)$. 
        The hardness factor has physical units inverse to those of the standard deviation. We set $h=10$~K$^{-1}$ for $T$ and $h=10$~psu$^{-1}$ for $S_{\rm u}$, and $h=100$~m$^{-1}$ for $\eta_{\rm u}$. The factor 1 in the argument of the logarithmic function ensures that \mbox{$\sigma_{\textrm{m}} = w_{\textrm{c}} \sigma_{\textrm{c}}$} when \mbox{$\sigma_{\textrm{p}} = 0$} and \mbox{$\sigma_{\textrm{m}} = w_{\textrm{p}} \sigma_{\textrm{p}}$} when \mbox{$\sigma_{\textrm{c}} = 0$}, although, in practice, $\sigma_{\textrm{c}}$ and $\sigma_{\textrm{p}}$ never attain zero values since they are bounded below with limiting values. 
        
        We refer to the hybridized value ($\sigma_\textrm{m}$) as the {\it modelled} background-error standard deviation. For the parameterized standard deviations, we set \mbox{$w_\textrm{p}=1$} for all variables. For the climatological standard deviations, we set \mbox{$w_\textrm{c}=1$} for $T$ and $S_{\rm u}$, and \mbox{$w_\textrm{c}=0.5$} for $\eta_{\rm u}$. The climatological standard deviations are down-weighted for $\eta_{\rm u}$ as we found that applying them at the full magnitude makes the assimilation system draw too strongly to the altimeter observations, which results in degraded performance. Subsequently, the modelled standard deviations, $\sigma_\textrm{m}$, are combined, also using the smooth maximum function, with ``errors of the day'', $\sigma_\textrm{e}$, estimated from the forecast ensemble:
        \begin{equation}
            \label{eq:smoothmax_h}
            \sigma_{\textrm{h}} = \frac{1}{h} \log \left( e^{h w_{{\textrm{m}}} \sigma_\textrm{m}} + e^{h w_\textrm{e} \sigma_\textrm{e}} - 1 \right)
        \end{equation}
        where $w_\textrm{m}$ and $w_\textrm{e}$ are weighting factors  for the modelled and ensemble-
        derived standard deviations, respectively. We refer to $\sigma_\textrm{h}$ as the hybrid (modelled--flow-dependent) standard deviations. Similar to the hybridization of the climatological component in Equation~(\ref{eq:smoothmax_m}), we set \mbox{$w_\textrm{m} = w_\textrm{e} =1$} for $T$ and $S_{\textrm{u}}$,  \mbox{$w_{\textrm{m}} = 1$} and \mbox{$w_{\textrm{e}} = 0.5$} for $\eta_{\textrm{u}}$, and used the same values of $h$.

        \begin{figure}
            \centering
            \begin{tabular}{c c}
                \includegraphics[width=0.35\linewidth]{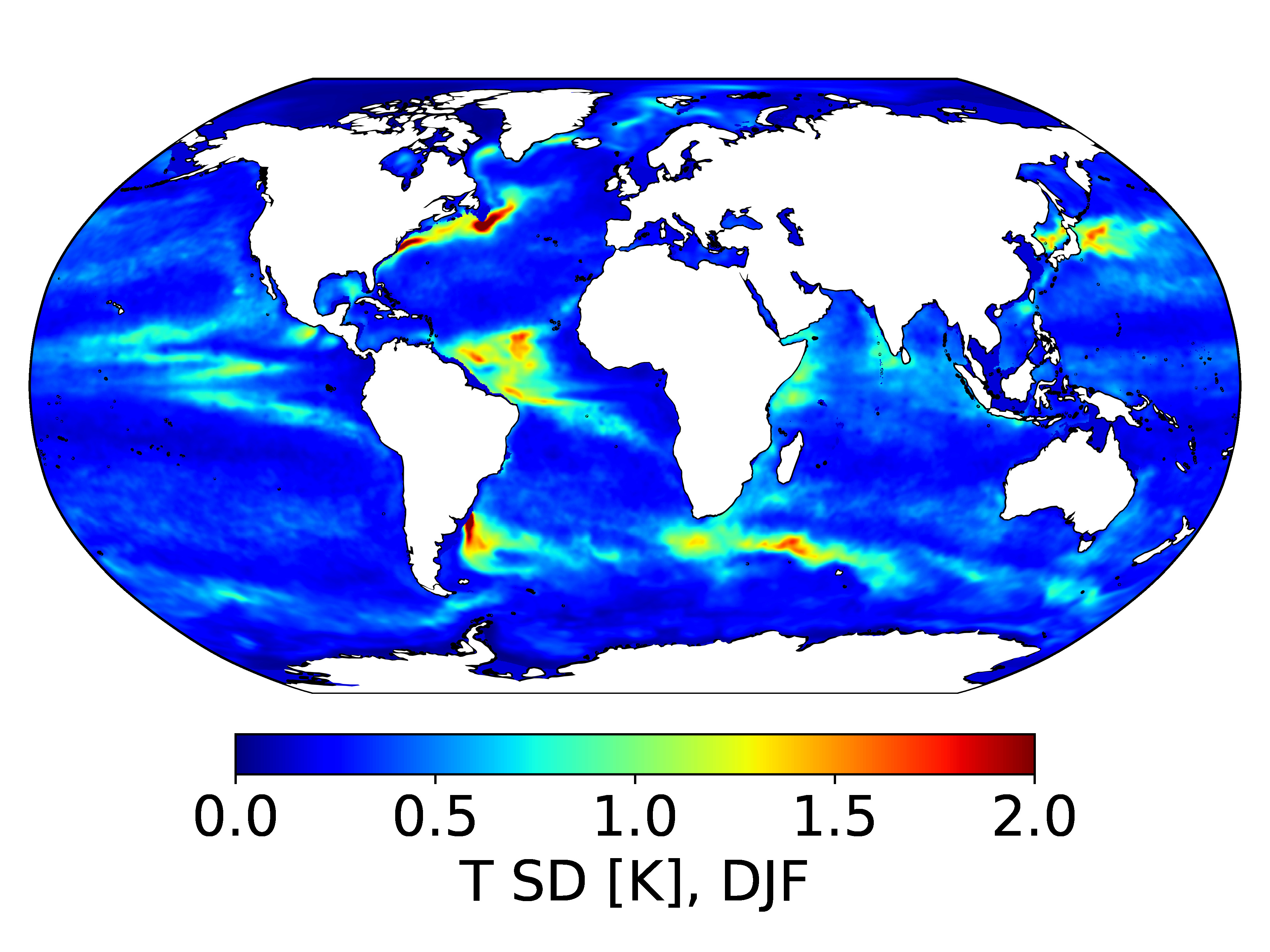}                 & \includegraphics[width=0.35\linewidth]{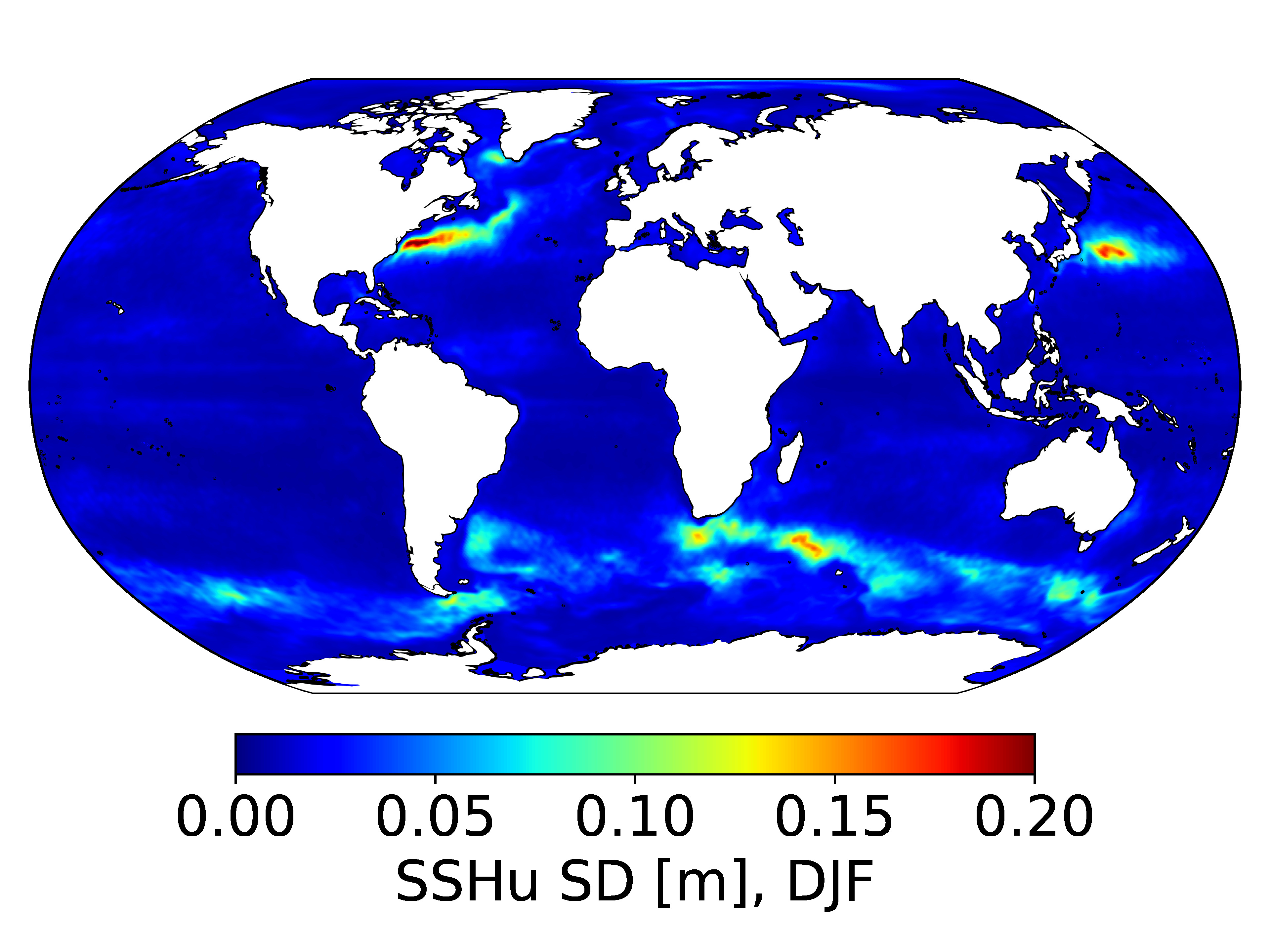} \\
                \includegraphics[width=0.35\linewidth]{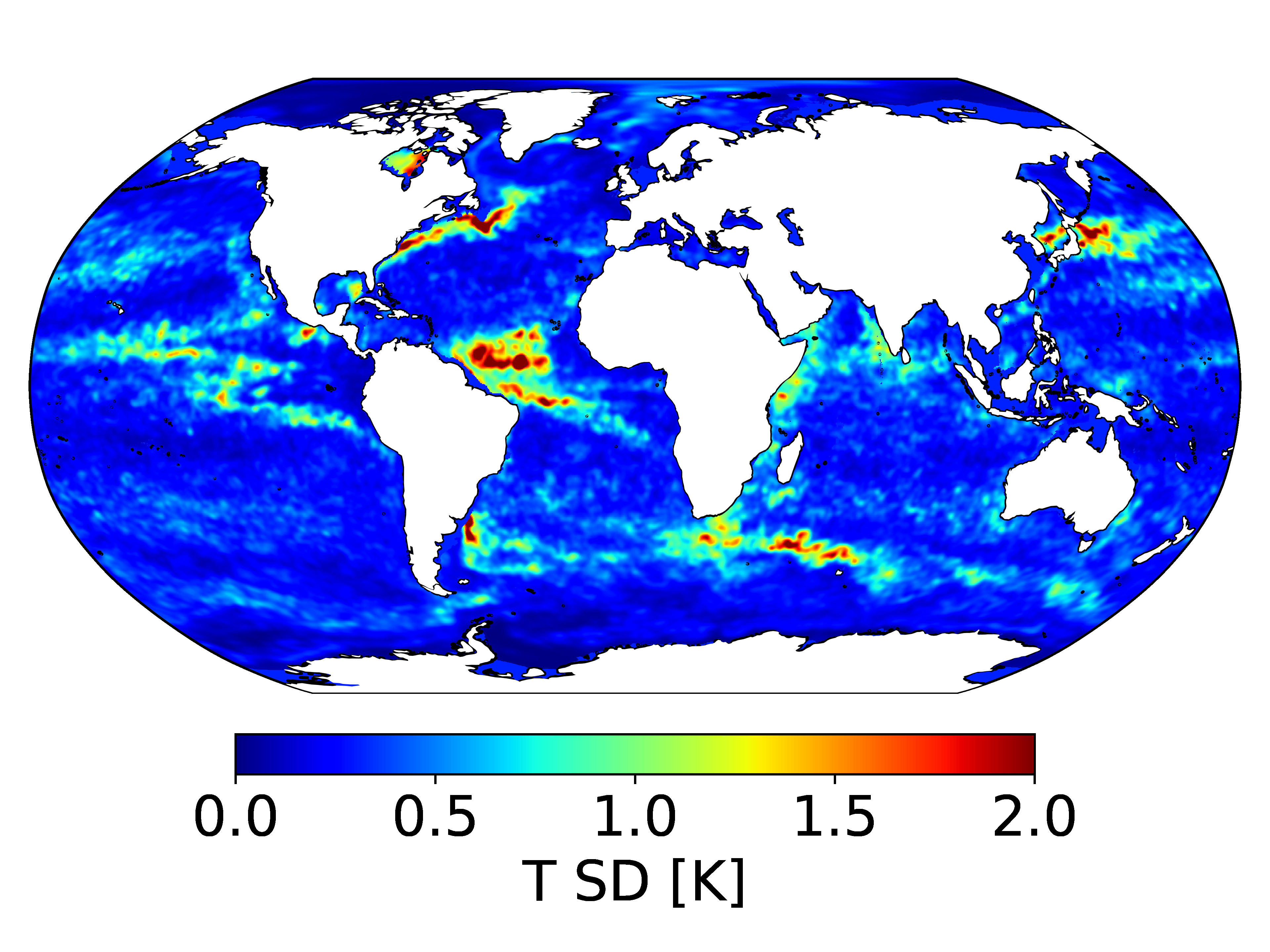}     & \includegraphics[width=0.35\linewidth]{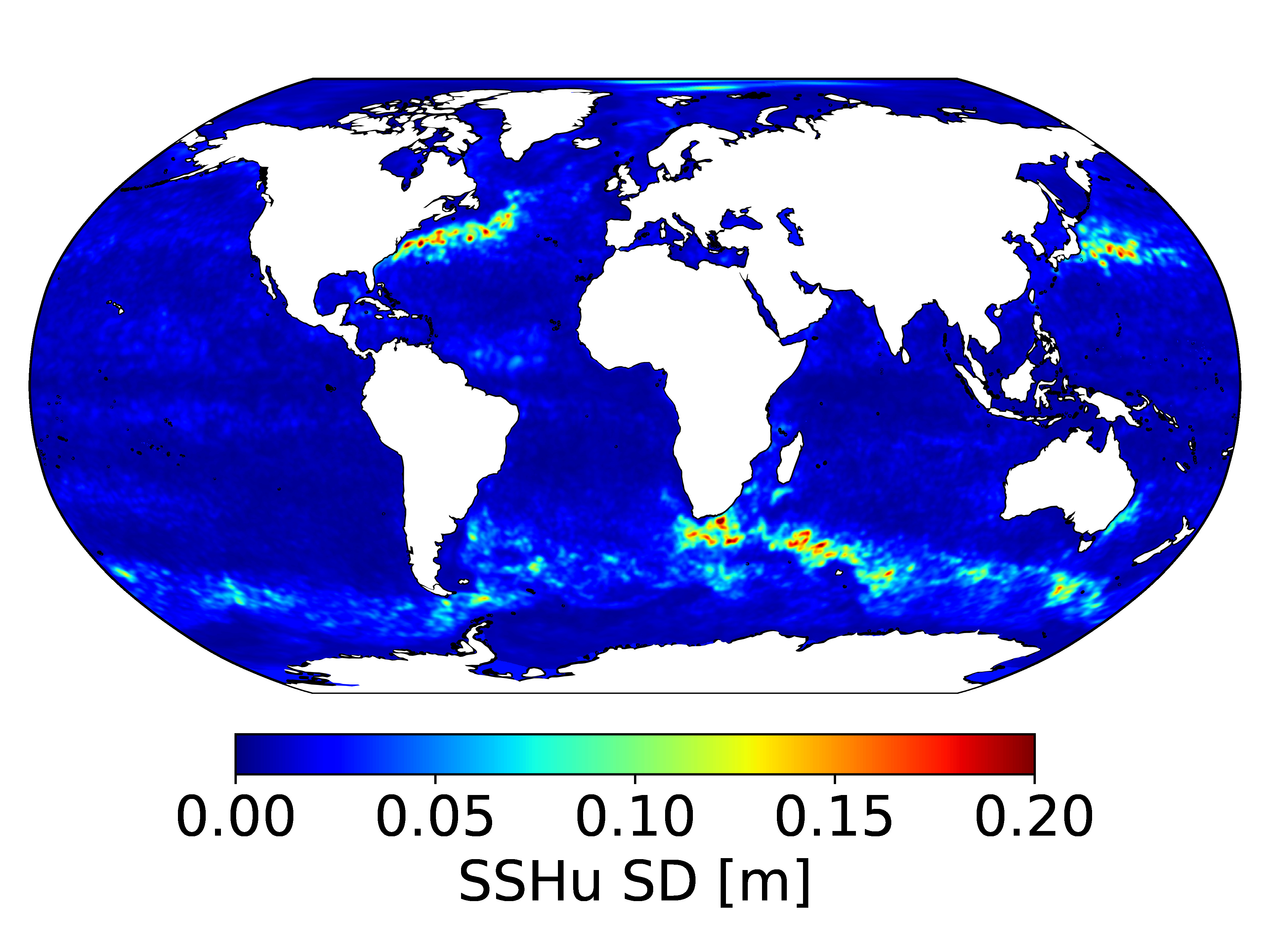} \\
                \includegraphics[width=0.35\linewidth]{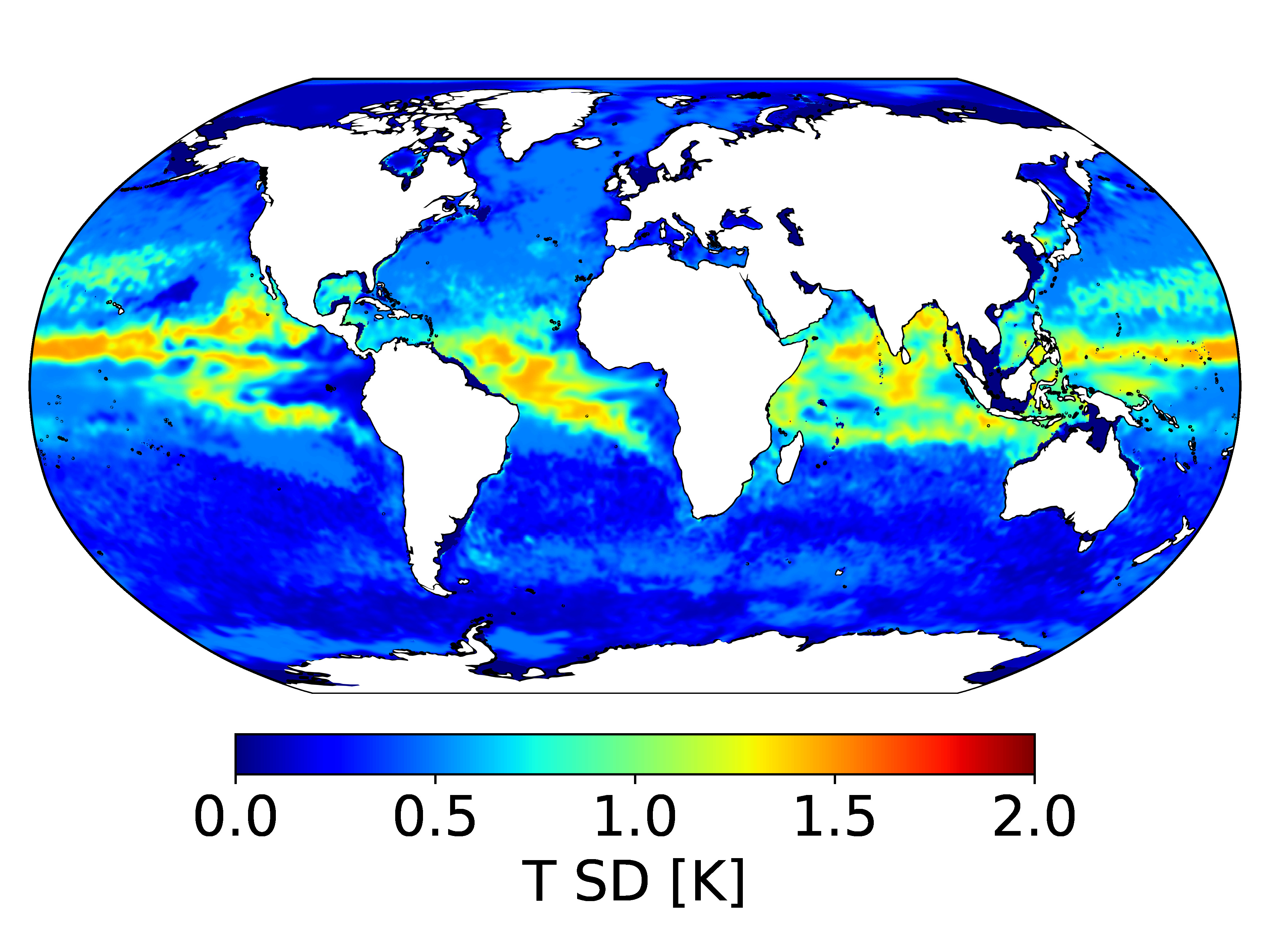}   & \includegraphics[width=0.35\linewidth]{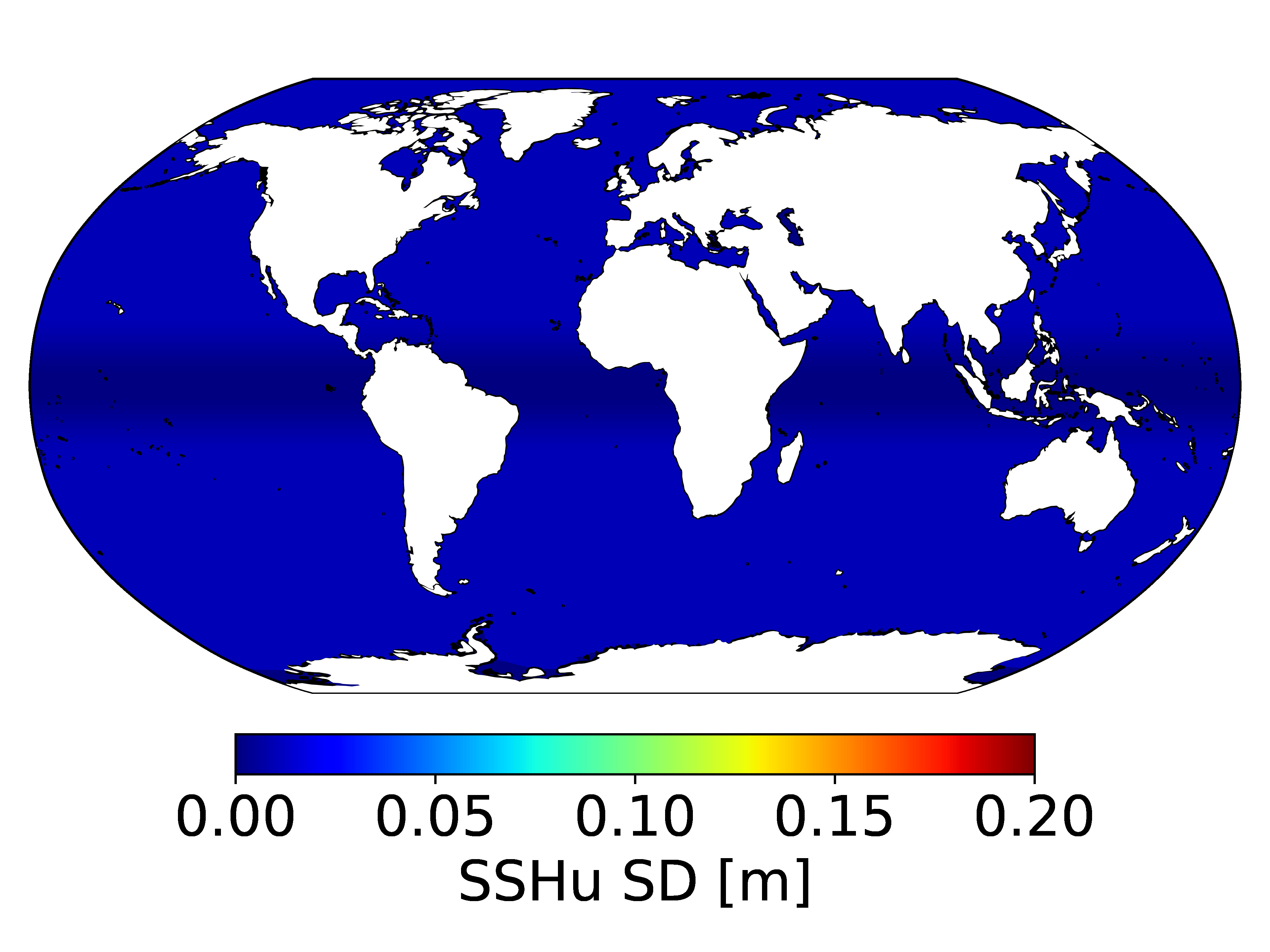} \\
                \includegraphics[width=0.35\linewidth]{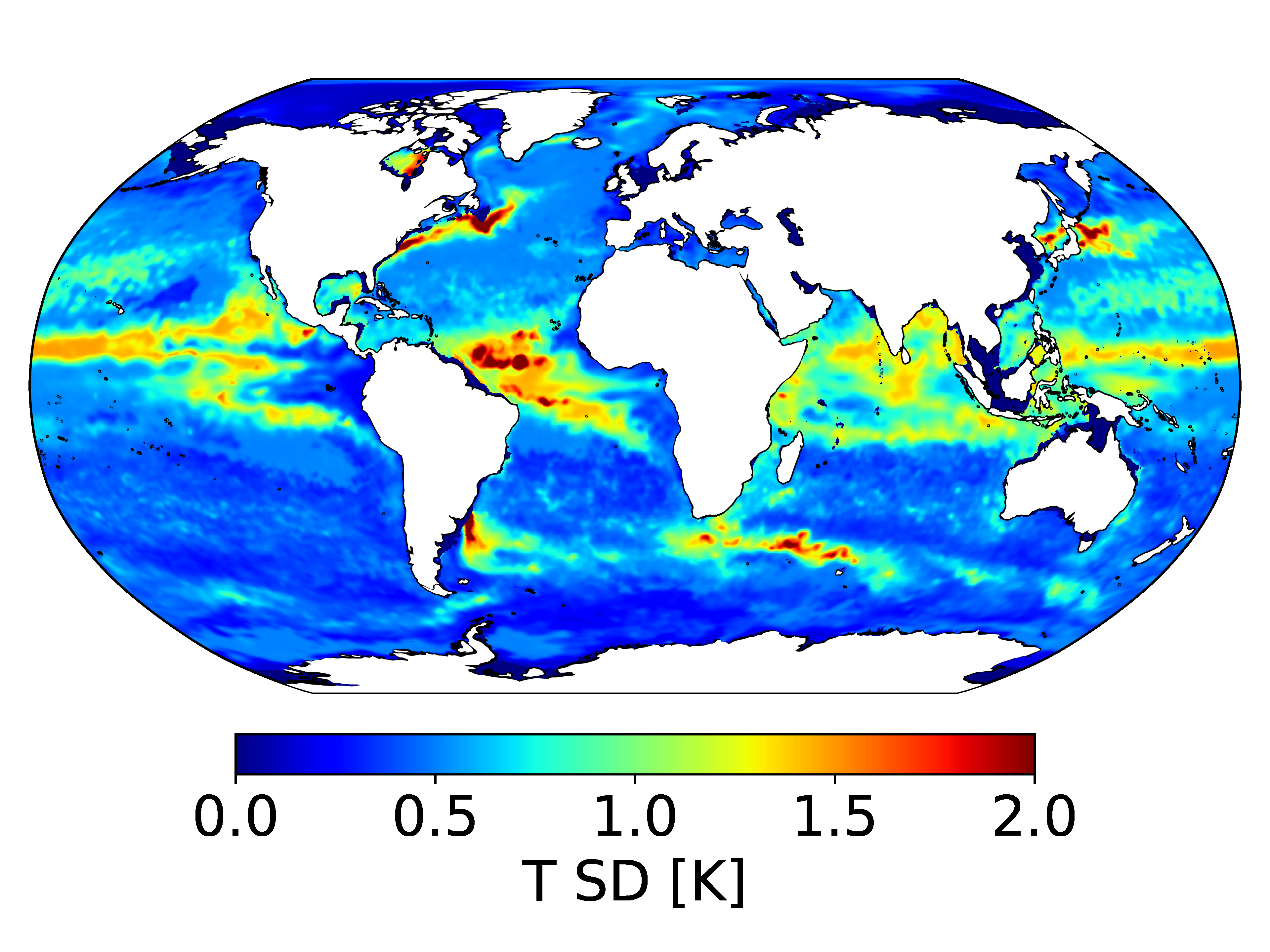}         & \includegraphics[width=0.35\linewidth]{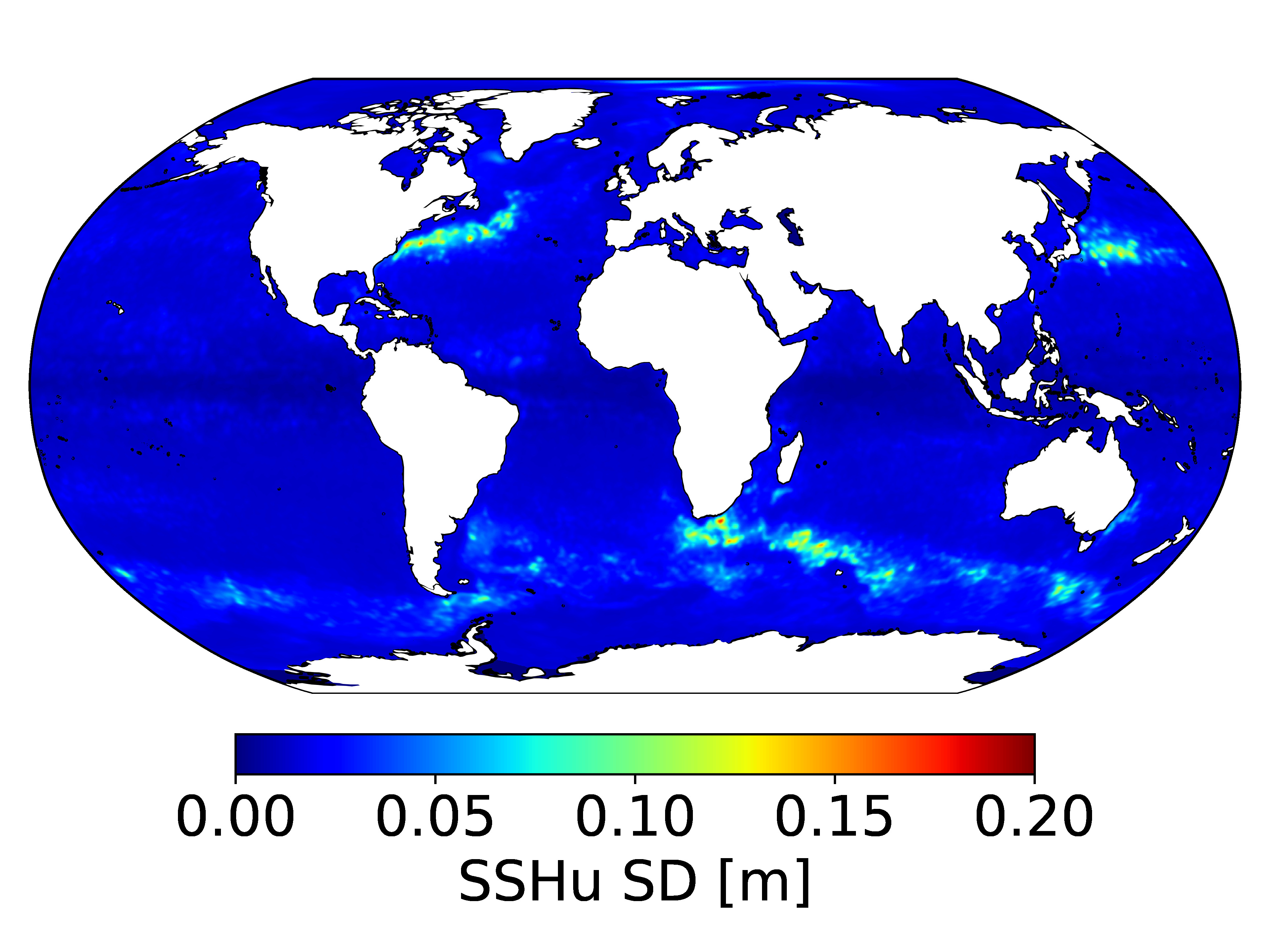} 
            \end{tabular}
            \caption{Rows from the top: climatological (estimated from ensemble perturbations for months December-January-February (DJF) in the period 2010--2015), ``errors of the day'', parameterized and hybrid standard deviations (SD) for temperature at 97~m depth and the unbalanced component of sea surface height (SSHu) (right column) for the cycle starting on 04/01/2017.}
            \label{fig:parm-clim-ens-sdv}
        \end{figure}

        As described above, the standard deviations specified in the new system are obtained by combining, using \Cref{eq:smoothmax_h}, the legacy ORAS5 parameterized standard deviations with climatological standard deviations and ``errors of the day'' diagnosed from the ensemble.
        This is illustrated \Cref{fig:parm-clim-ens-sdv} for a particular cycle, valid on January $4^\textrm{th}$, $2017$, for temperature, in level~24 of the ocean model, and for unbalanced SSH. The climatological background-error standard deviations are illustrated in the top row for the boreal winter (December-January-February). Not surprisingly, the errors are largest in eddy-active regions  and rather small otherwise. This is well visible for the unbalanced SSH standard deviations, which appear to be close to zero apart from the WBC and ACC regions. Seasonal variability, not shown in the figure, is generally small. It can be best observed for the Gulf Stream and Kuroshio Current regions, where the errors are larger in the boreal winter and relatively smaller in the boreal summer. The standard deviations corresponding to the ``errors of the day'', illustrated in the second row, show sharp flow-dependent features in eddy-active regions. Kelvin wave-like structures are visible in the tropical Pacific for temperature. On the other hand, elsewhere the standard deviations are suppressed compared to the parameterized values. This is particularly visible for unbalanced SSH. It is interesting to contrast the ``errors of the day'' with the climatological values. The structures are, as expected, much smoother. The location of the largest standard deviations is unsurprisingly very similar and confined to eddy-active regions The parameterized standard deviations, illustrated in the thrid row of the figure, show a limited degree of flow dependency in the extra-tropics in the case of temperature and are constant for unbalanced SSH. It should be noted that in the case of unbalanced SSH, the standard deviations are smoothly damped to zero in the tropics. Finally, the last row shows the combined hybrid standard deviations. Owing to the properties of the smooth maximum function, the structure resembles that of sharpened climatological standard deviations with the background magnitude outside of eddy-active regions corresponding to that of the parameterized values.
        
\begin{figure}
            \centering
            \begin{tabular}{c c}
            \includegraphics[width=0.45\linewidth]{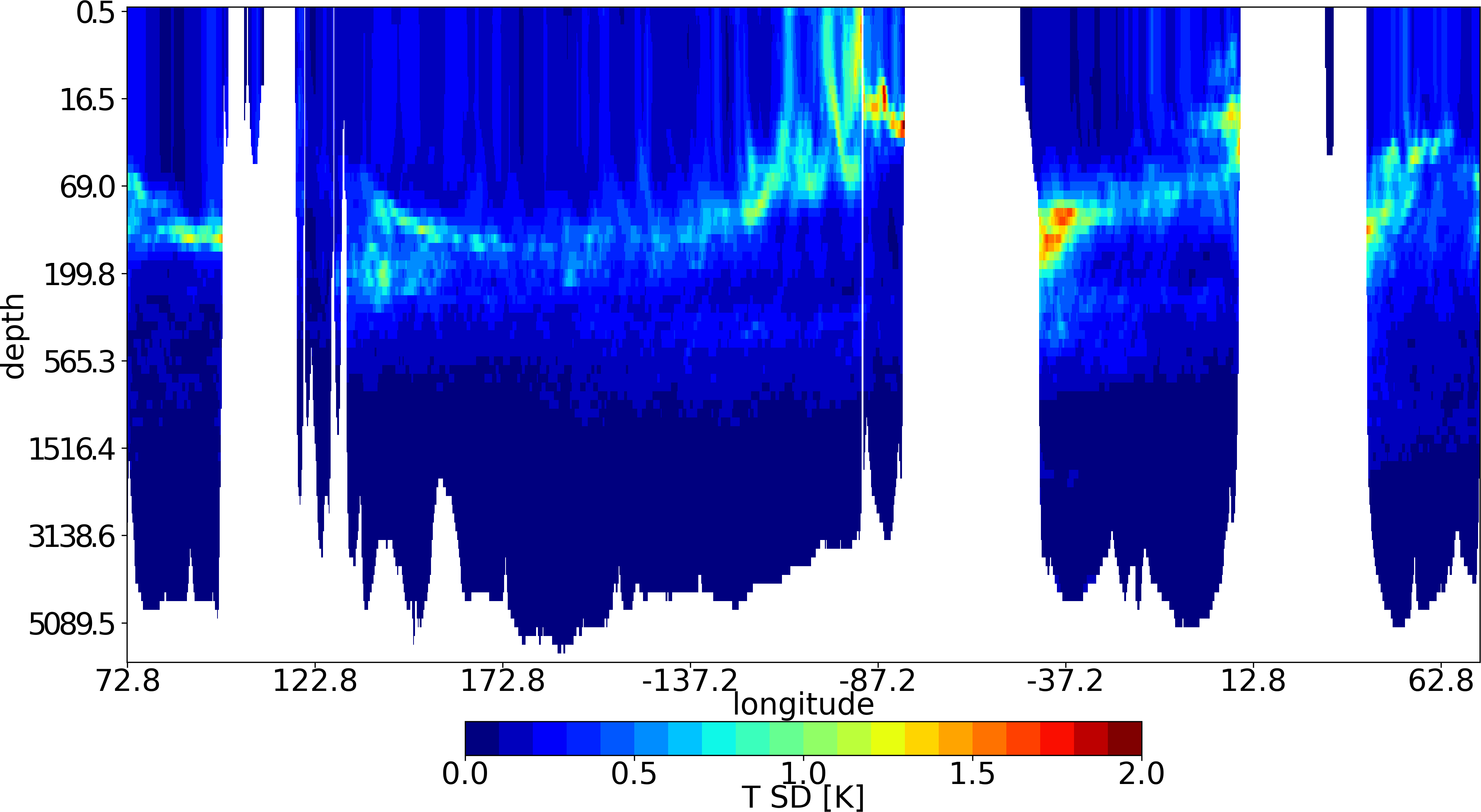} & \includegraphics[width=0.45\linewidth]{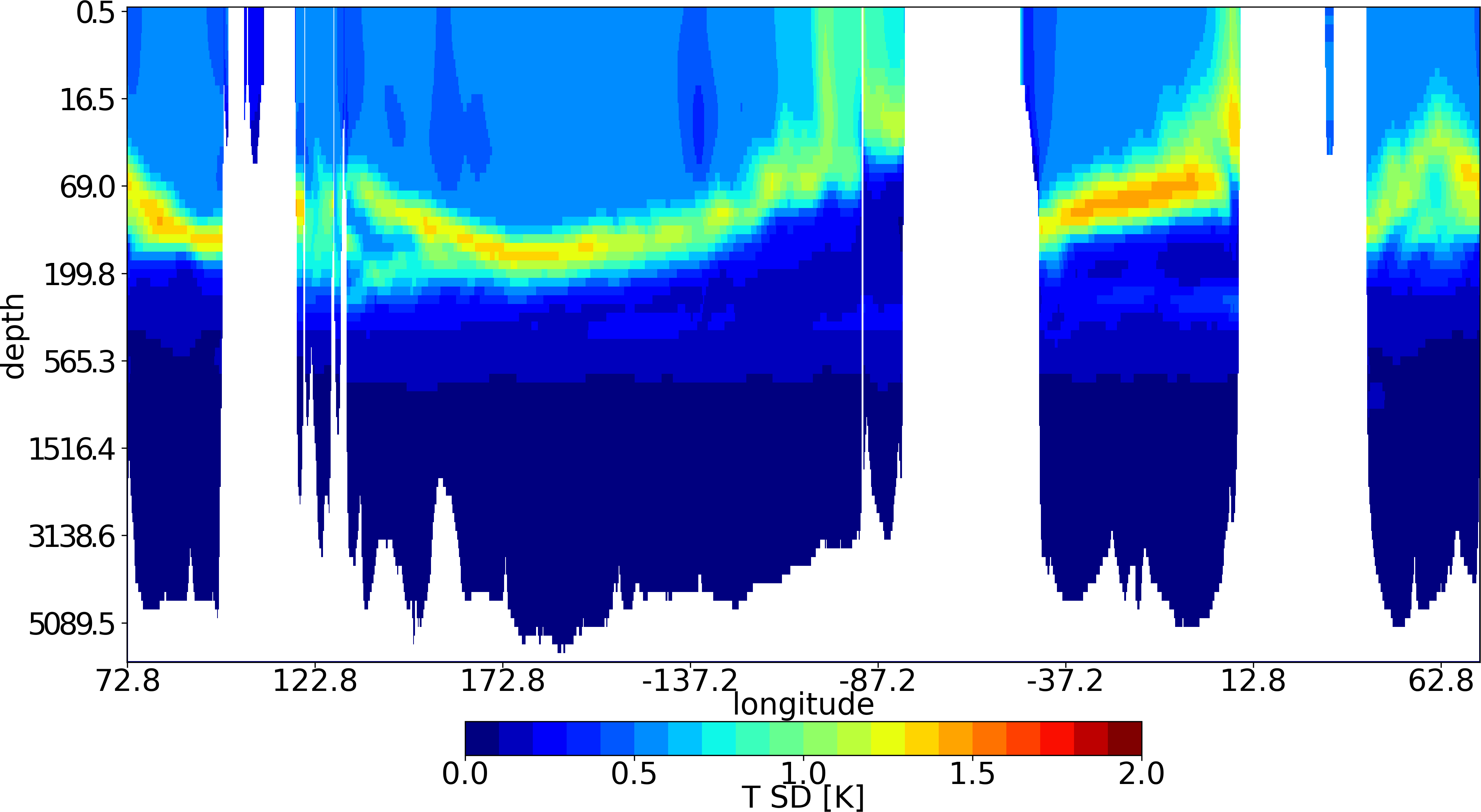} \\
            \includegraphics[width=0.45\linewidth]{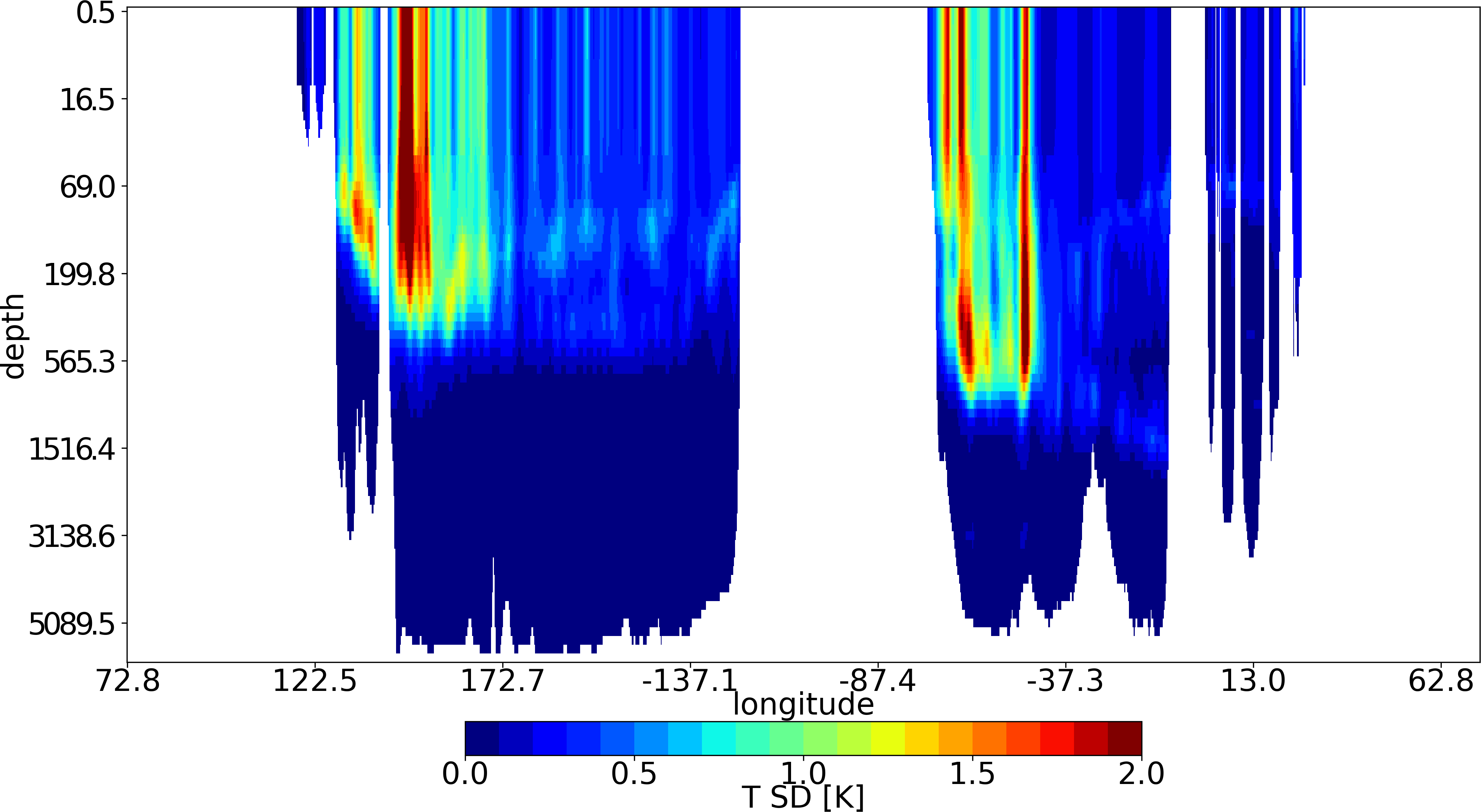} & \includegraphics[width=0.45\linewidth]{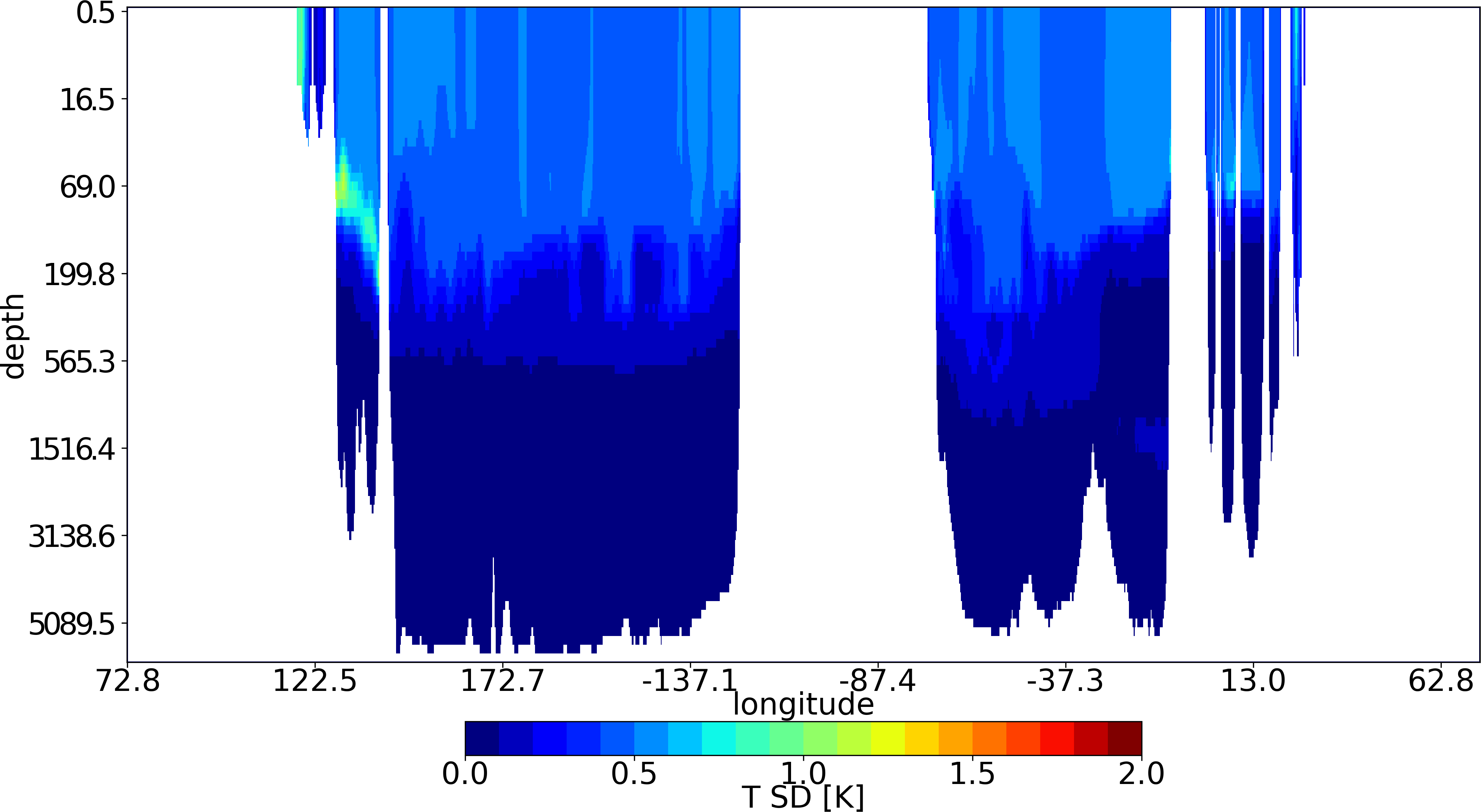} 
            \end{tabular}
            \caption{Vertical section of the background-error standard deviations (SD) for temperature estimated from the ensemble (left panel) and from the ORAS5 parameterization (right panel) at the Equator (top row) and at a constant latitude of $40^\circ$N (bottom row), valid on 04/01/2017.}
            \label{fig:clim-sdv-t-vert-40N}
        \end{figure}

        \Cref{fig:clim-sdv-t-vert-40N} shows the vertical structure of the ensemble and parameterized background-error standard deviations at the Equator (top row) and at the constant latitude of $40^\circ$N (bottom row), valid on January~$4^\textrm{th}$, 2017. Compared to the parameterized values, the ensemble-derived parameters show similar vertical structure at the Equator, but the parameterized values are substantially larger in the mixed layer and at the level of the thermocline. At the constant latitude of $40^\circ$N, the ensemble-derived parameters show richer vertical and horizontal structure, and are larger in the mixed layer in the Gulf Stream and Kuroshio Current regions and smaller elsewhere apart from the deep ocean where both are close to zero. 

        Examining the new hybrid background-error standard deviations, it is clear that the background forecast is given less confidence in the eddy-active regions in the new system and as a consequence is allowed to draw more to observations there. Such behaviour is desirable in order to improve the accuracy of the initial state  where the uncertainty in the background state is highest.

        \subsubsection{Hybrid diffusion tensor} \label{subsec:bhybclim-ten}

        As described in \Cref{subsec:b}, the implicit diffusion operator used to model the correlation matrix requires normalization. Computing accurate normalization factors for a 3D implicit diffusion operator is costly, which precludes us from using a fully flow-dependent 3D correlation model. When the horizontal and vertical correlations are modelled separately as in Equation~(\ref{eq:V_alpha}), \cite{weaver-2021} showed that the normalization matrix can be well approximated by a product of a normalization matrix for the horizontal diffusion component and a normalization matrix for the vertical diffusion component, where each can be estimated separately using randomisation or exactly in the case of the latter. This approximation allows us to decouple the specification of the horizontal and vertical diffusion tensor components. While the cost of computing the normalization factors of the horizontal component is high and can be considered not affordable in a cycling system, the cost of computing the vertical normalization factors is low. For this reason, the new system uses a climatological horizontal diffusion tensor and a fully flow-dependent vertical diffusion tensor. 
        
\begin{figure}
            \centering
            \begin{tabular}{c c}
            \includegraphics[width=0.35\linewidth]{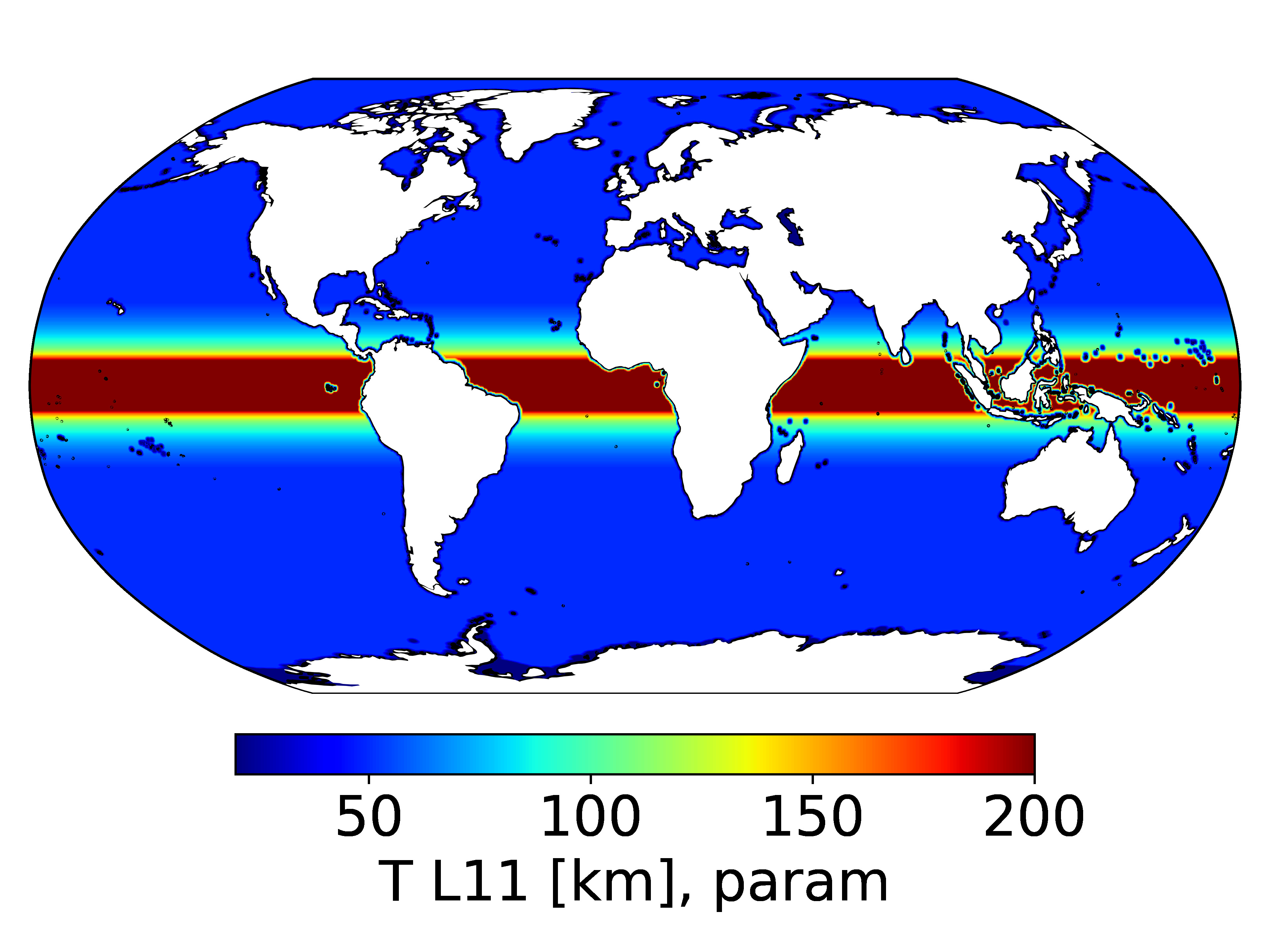}     & \includegraphics[width=0.35\linewidth]{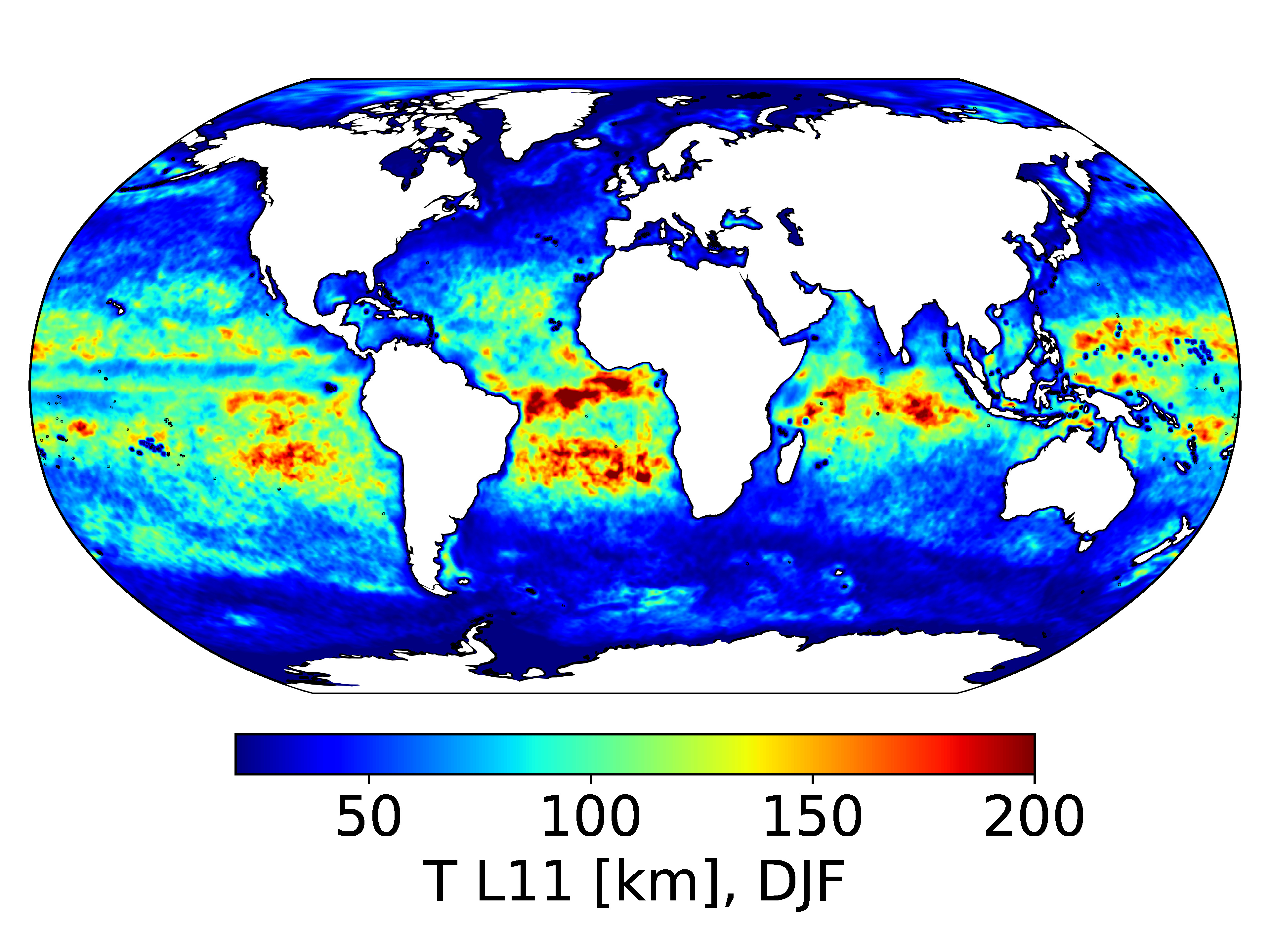} \\
            \includegraphics[width=0.35\linewidth]{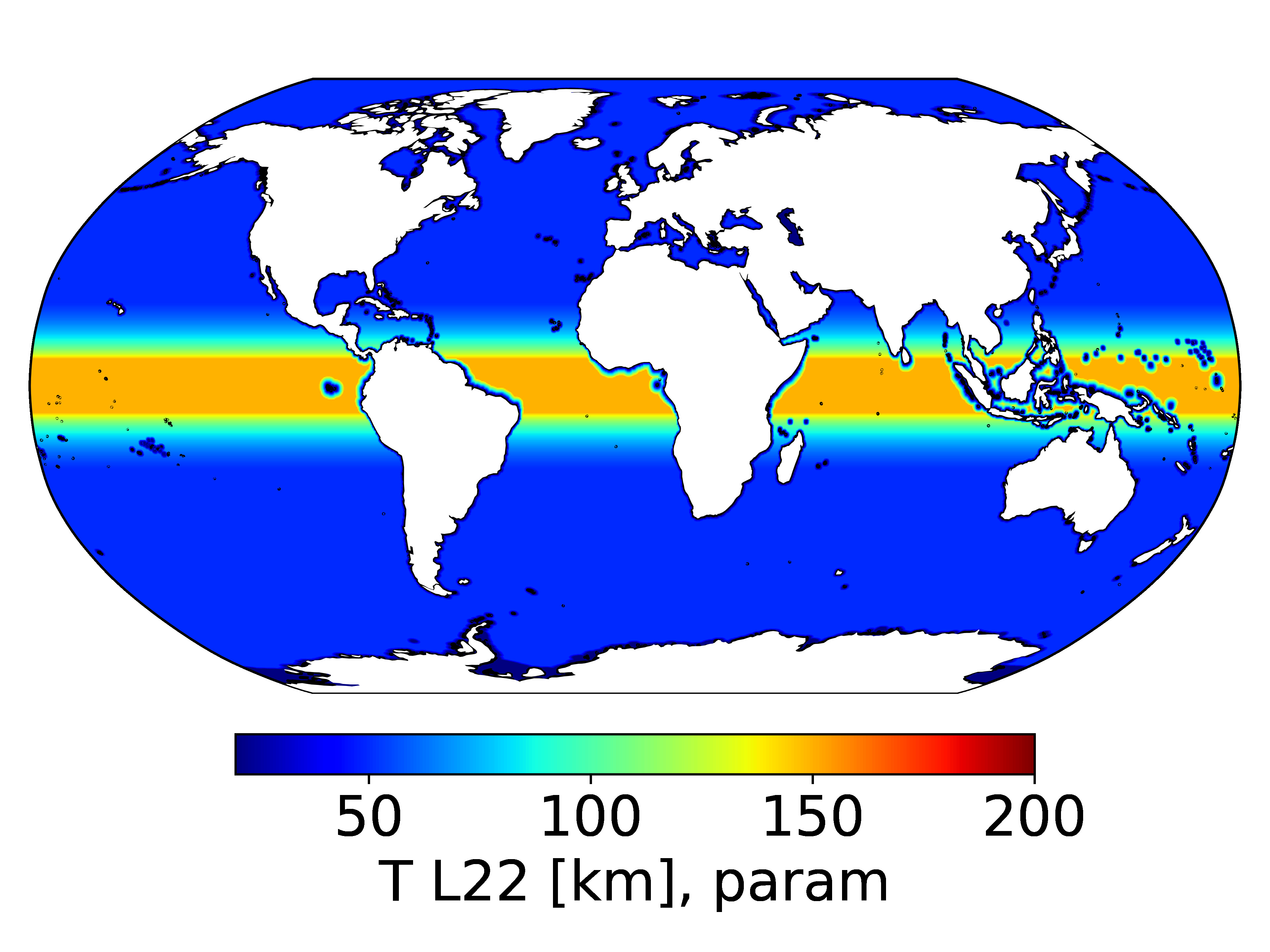}     & \includegraphics[width=0.35\linewidth]{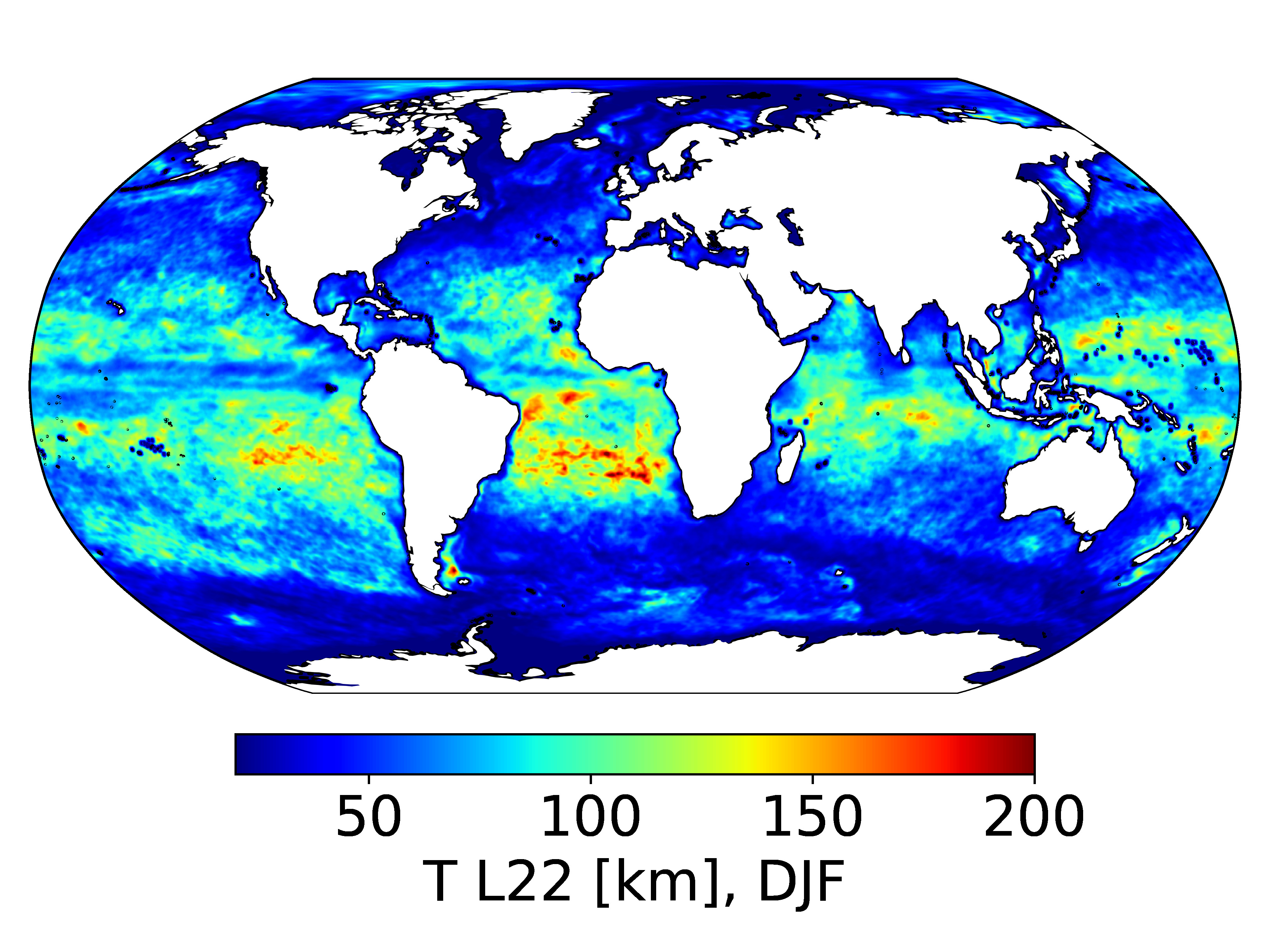} \\
            \end{tabular}
            \caption{Temperature meridional (L11, top row) and zonal (L22, bottom row) horizontal background-error correlation length-scales at 1~m depth. Left: ORAS5 parameterized length-scales. Right: climatological length-scales estimated from ensemble perturbations for months December-January-February (DJF) in the period 2010--2015.}
            \label{fig:parm-clim-ten-t}
        \end{figure}

        \Cref{fig:parm-clim-ten-t} shows a comparison of the zonal and meridional components of the horizontal correlation length-scales for temperature at $1$~m depth between ORAS5 in the left column and the new system in the right column. In ORAS5, the horizontal correlation length-scales, which are proportional to the square-root of the diagonal elements of the horizontal diffusion tensor, were specified for temperature using a parameterization based on the Rossby radius. The correlation length-scales are largest in the tropics and decrease with increasing latitude. Zonal correlation length-scales are larger than meridional correlation length-scales, reflecting well known anisotropy related to equatorial dynamics. The anisotropy is also clearly visible in the climatological length-scales. Similarly, the length-scales are largest in the tropics, albeit shorter than in ORAS5, and smaller in the extra-tropics, in particular in the WBC and ACC regions. In the tropics, the climatological length-scales show clear minima associated with the location of Tropical Instability Waves. It is also interesting to observe that larger correlation length-scales are diagnosed in the Southern Hemisphere in the boreal winter, which is not the case in boreal summer (not shown).
        
\begin{figure}
            \centering
            \begin{tabular}{c c}
            \includegraphics[width=0.35\linewidth]{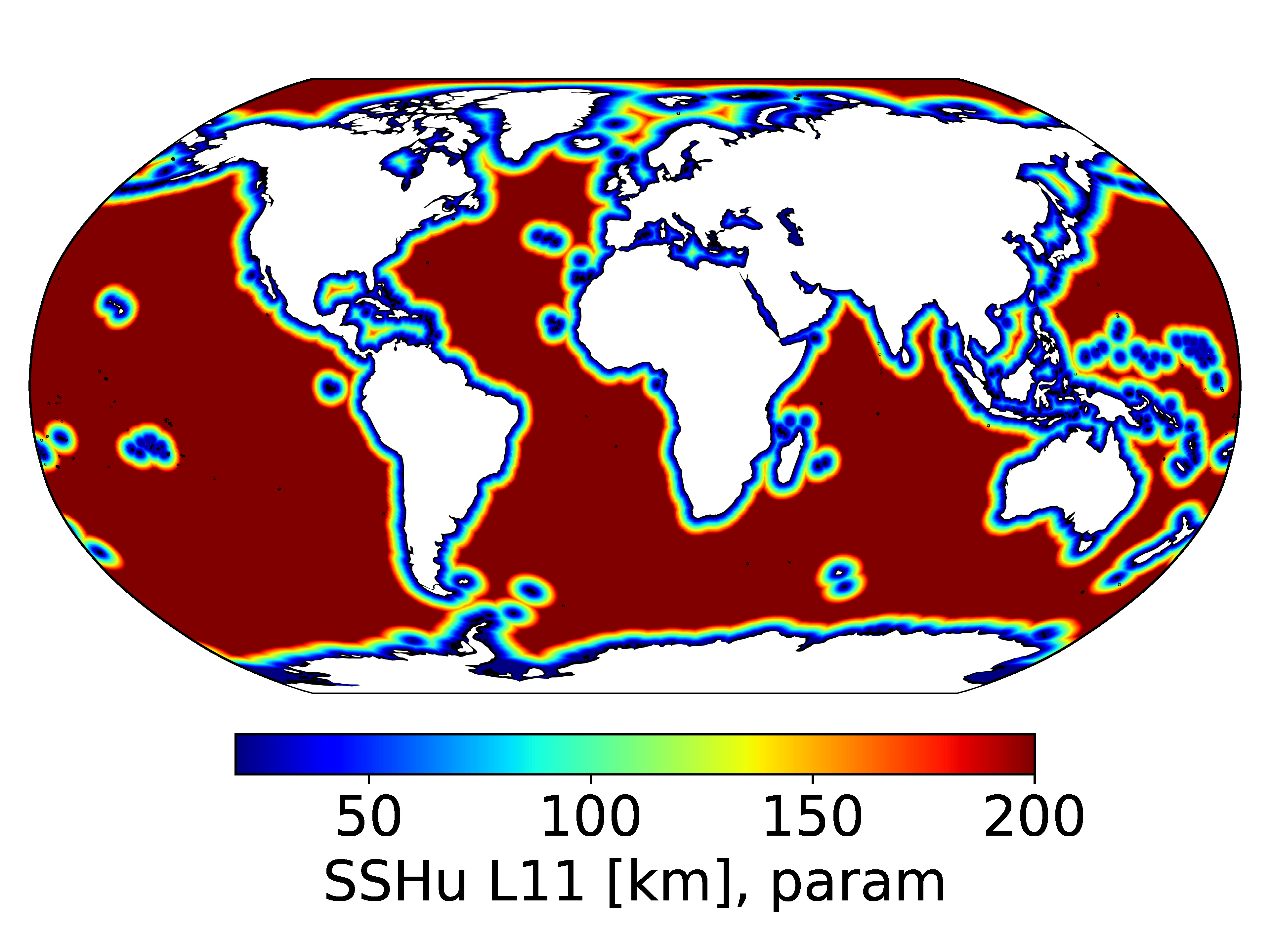}     & \includegraphics[width=0.35\linewidth]{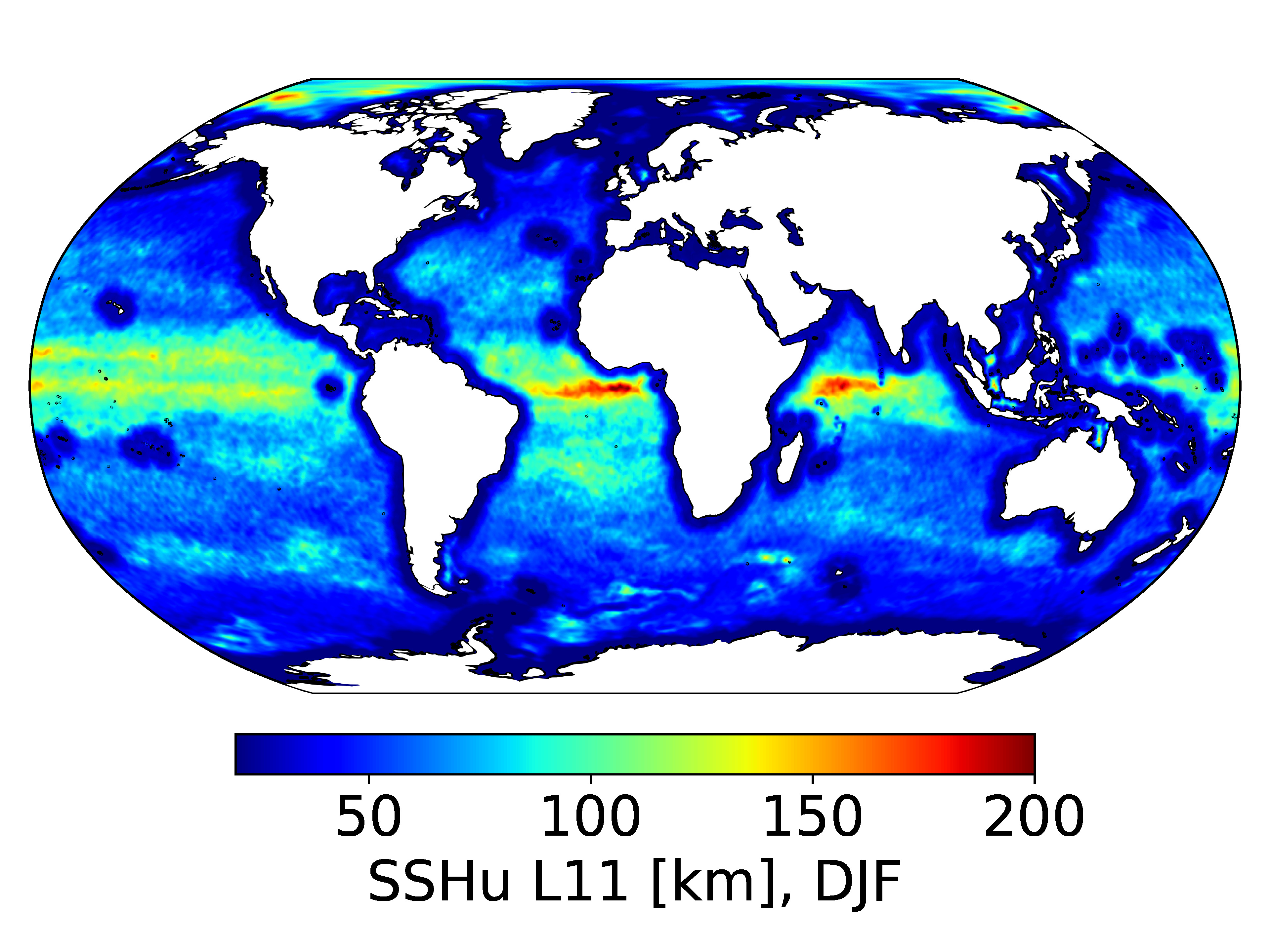} \\
            \includegraphics[width=0.35\linewidth]{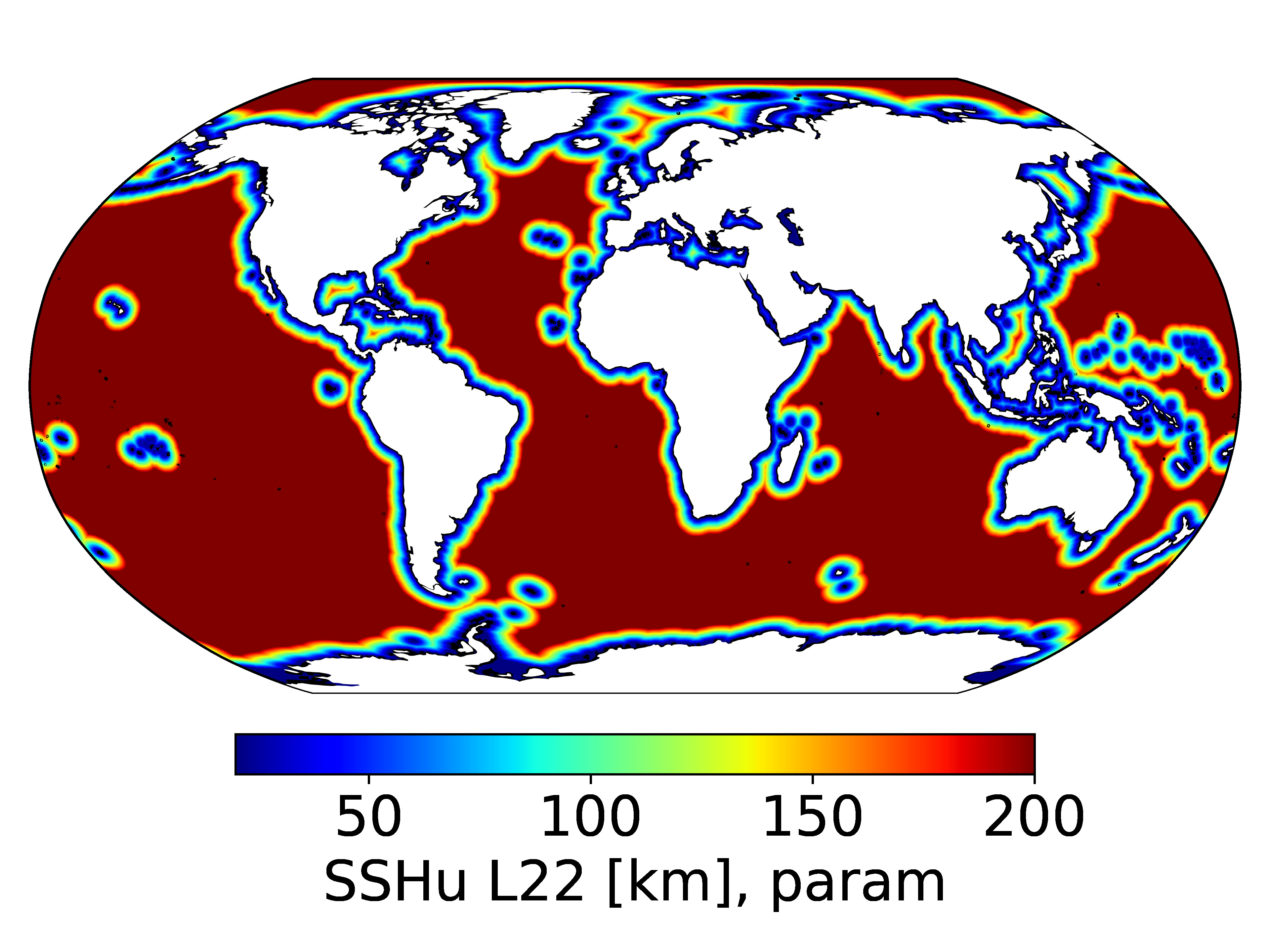}     & \includegraphics[width=0.35\linewidth]{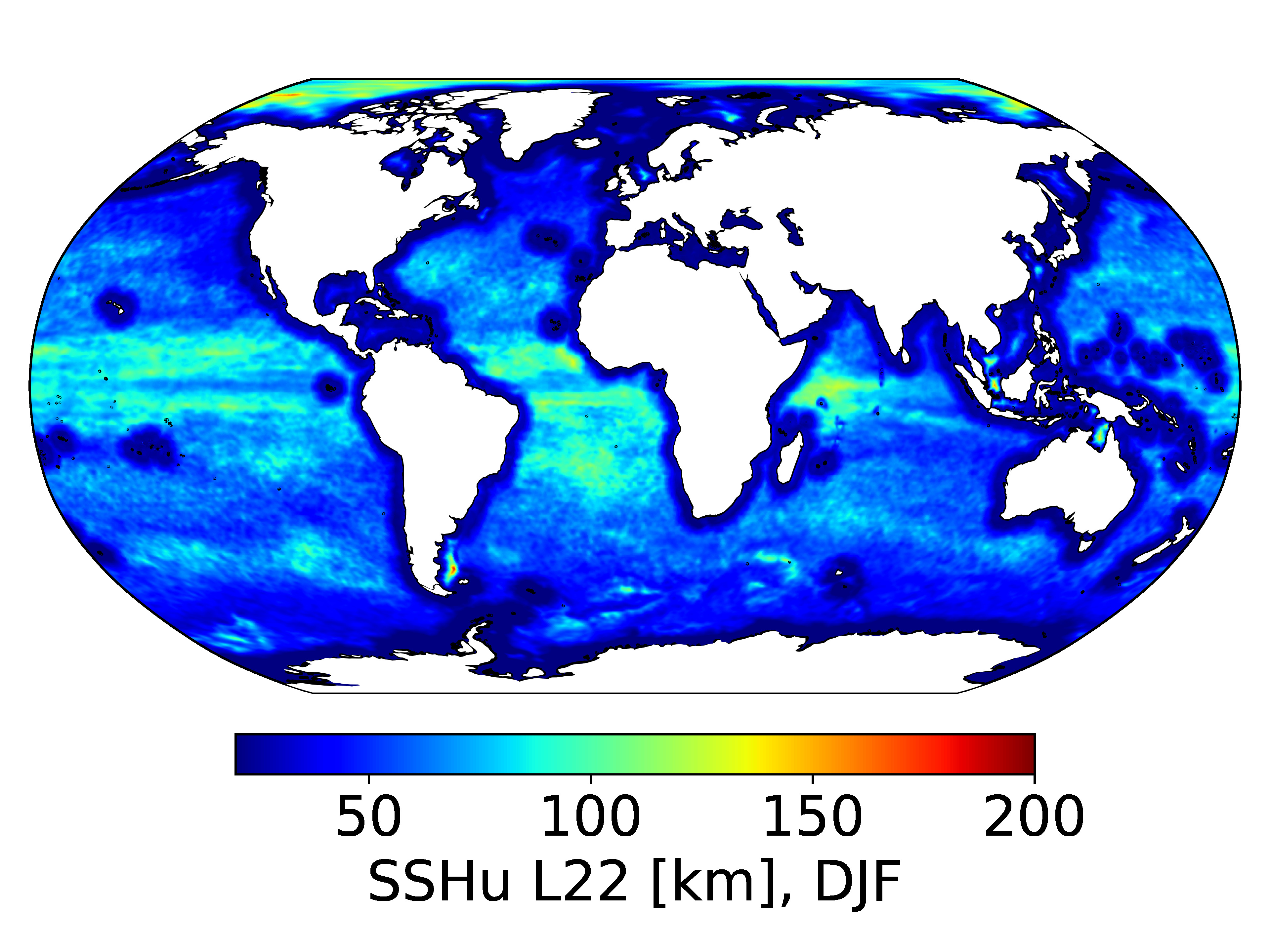} \\
            \end{tabular}
            \caption{Unbalanced component of SSH (SSHu) meridional (L11, top row) and zonal (L22, bottom row) horizontal background-error correlation length-scales. Left: ORAS5 parameterized length-scales. Right: climatological length-scales estimated from ensemble perturbations from months December-January-February (DJF) in the period 2010--2015.}
            \label{fig:parm-clim-ten-ssh}
        \end{figure}

        \Cref{fig:parm-clim-ten-ssh} shows the comparison of the zonal and meridional components of the horizontal correlation length-scales for unbalanced SSH. As can be seen in the left column of the figure, the zonal and meridional length-scales are equal and constant everywhere, except in the proximity of land boundaries where they are reduced. The climatological length-scales are significantly shorter and also reduced artificially in the proximity of land boundaries. Similarly, as for temperature, the zonal correlation length-scales are larger than the meridional correlation length-scales, reflecting the anisotropic nature of the dynamics in this region.  

\begin{figure}
           \centering
           \includegraphics[width=0.4\linewidth]{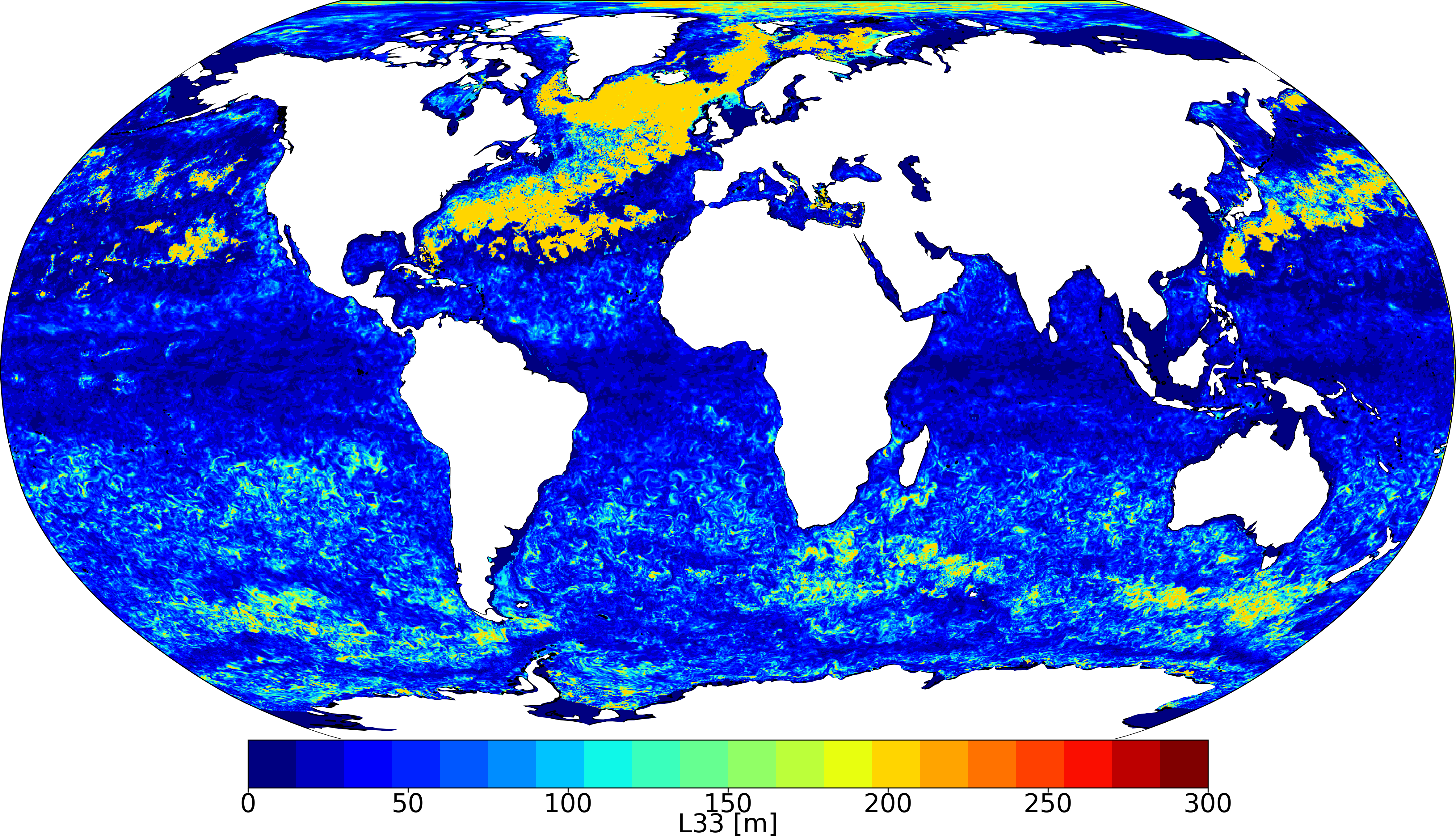} \\
           \vspace{1cm}
           \begin{tabular}{c c}
            \includegraphics[width=0.4\linewidth]{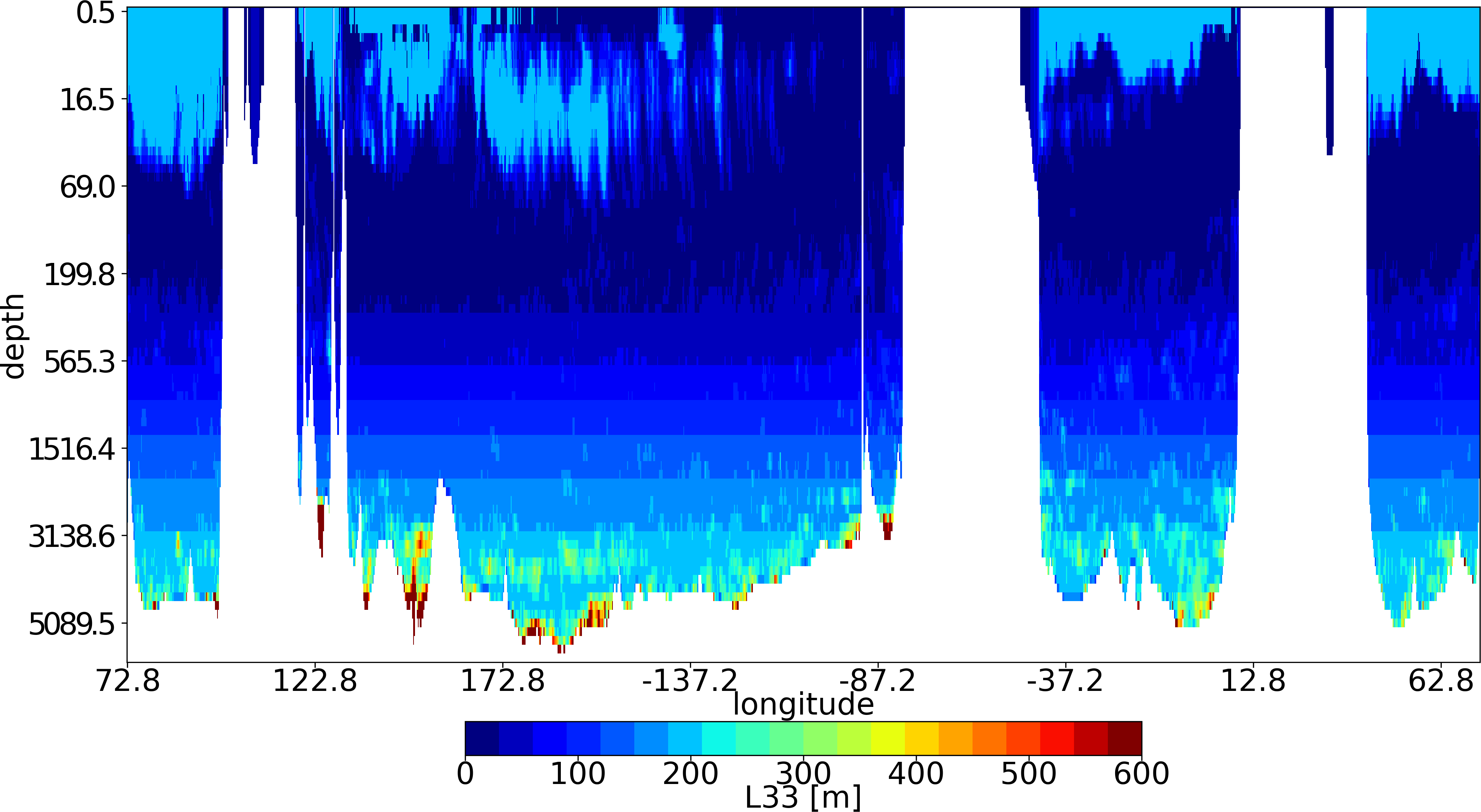} &
            \includegraphics[width=0.4\linewidth]{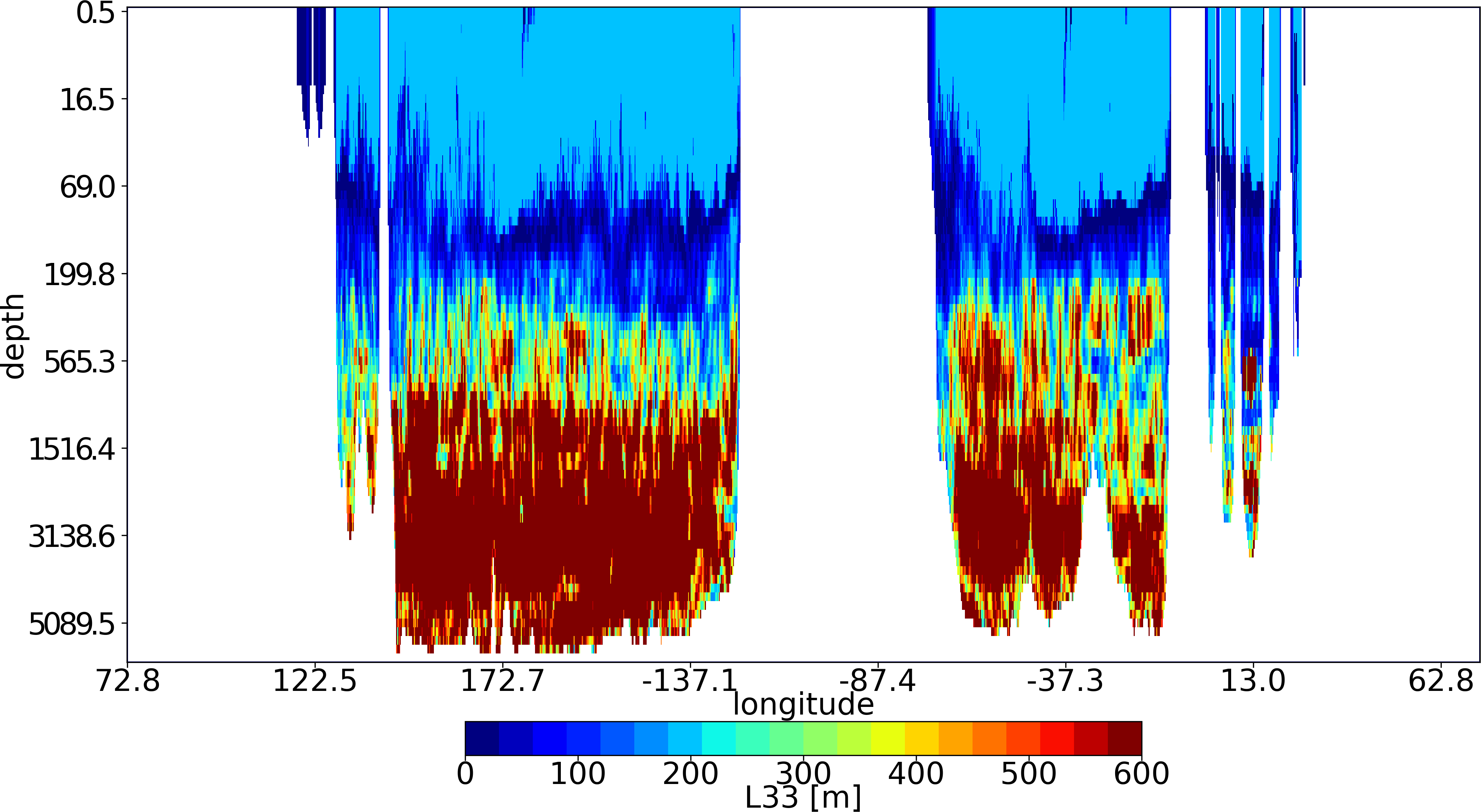}
            \end{tabular}
            \caption{Top row: vertical (L33) background-error correlation length-scales estimated from the ensemble for temperature at 97~m depth. Bottom row: vertical section at the Equator (left panel) and at the constant latitude of $40^\circ$N (right panel) of the vertical (L33) background-error correlation length-scales estimated from the ensemble. Plots are valid on 04/01/2017.}
            \label{fig:param-clim-L33}
        \end{figure}

        The vertical diffusion tensor coefficients are parameterized in ORAS5 such that the vertical correlation length-scales are proportional to the vertical grid-scale factors. The proportionality constant was set to be equal to one. The vertical correlation length-scales estimated from the ensemble are shown in \Cref{fig:param-clim-L33} for the assimilation cycle starting on January~$4^\textrm{th}$, 2017. The top row panel shows the vertical correlation length-scales at 97~m depth. The vertical length-scales near the surface are much larger in the Northern Hemisphere than in the Southern Hemisphere. This is consistent with the expectation that the errors are strongly correlated in the mixed layer, which is deeper in the boreal winter in the Northern Hemisphere. The bottom row panels show a vertical cross section of the vertical correlation length-scales at the Equator (left panel) and at the $40^{\circ}$N latitude (right panel). At $40^{\circ}$N latitude, the highly stratified thermocline region, where the vertical length-scales are markedly shorter, can be clearly distinguished from both the mixed layer and the deep ocean, where the length-scales are much larger. The parameterized formulation of the vertical diffusion tensor used in ORAS5, by construction, neither distinguishes the thermocline region, nor well represents the length-scales in the mixed layer and the deep ocean. In both cases, the vertical length-scales are much smaller than those estimated from the ensemble. In fact, the short vertical correlation length-scales in the mixed layer in the parameterized error covariances make the assimilation of SST ineffective, as will be illustrated in the next section.

\section{Results} \label{sec:results}

    In this section, we demonstrate the impact of the new, hybrid formulation of $\mathbf{B}$. We carry out a series of experiments in the observation-rich period spanning 2017--2021. The evaluation period is outside the period which we used to compute the climatology of the  parameters of $\mathbf{B}$. Before commenting on the results, we start by describing our baseline configuration in \Cref{subsec:boras5}. Rather than presenting the overall impact of the new, hybrid $\mathbf{B}$ formulation, we first show the impact of the hybrid variances and the hybrid tensor in isolation in \Cref{subsec:bhybvar-res} and \Cref{subsec:bhybten-res}, respectively. Finally, the impact of the complete hybrid $\mathbf{B}$ formulation is presented in \Cref{subsec:bhybtenvar-res}. Our evaluation metric focuses on analysing the impact on Root Mean Squared (RMS) errors of the background forecasts relative to the observations (before assimilation). We consider the new system to be an improvement over  the baseline configuration whenever the background errors are reduced relative to those from the baseline configuration. We do not attempt in this article to analyse the impact on climate trends. This topic will be addressed in a separate article.

    \subsection{Baseline configuration} \label{subsec:boras5}
    
    Our baseline experiment for evaluating the impact of the new, hybrid $\mathbf{B}$ formulation employs the strictly parameterized formulation of $\mathbf{B}$ that was used for ORAS5. The assimilation window is $5$~days like in ORAS5. The assimilated observations include information on temperature and salinity from {\it in situ} measurements (Argo floats, moored buoys, ships, marine mammals, and gliders), sea level anomalies (SLA) from altimeters, and sea ice concentration from L3 satellite-based products. Unlike in ORAS5, the L4 OSTIA SST data are assimilated in NEMOVAR. All other aspects of the system, including the model bias correction scheme and the mean dynamic topography used for referencing SLA observations are those developed for ORAS6. Our goal is to evaluate the impact of the hybrid $\mathbf{B}$ formulation with respect to its parameterized predecessor.

    \subsection{Impact of the hybrid variances} \label{subsec:bhybvar-res}

\begin{figure}
            \centering
            \begin{tabular}{c c}
            \includegraphics[width=0.35\linewidth]{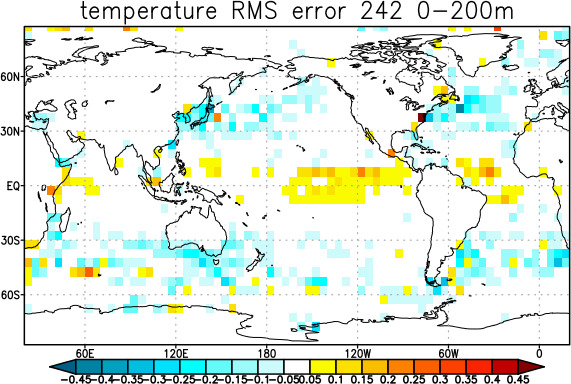}  &
            \includegraphics[width=0.35\linewidth]{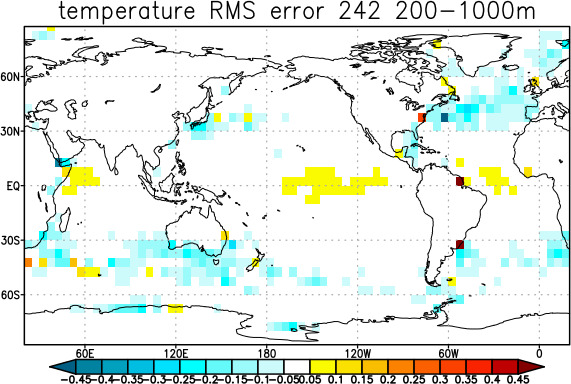}            \vspace{0.25cm} \\
            \includegraphics[width=0.35\linewidth]{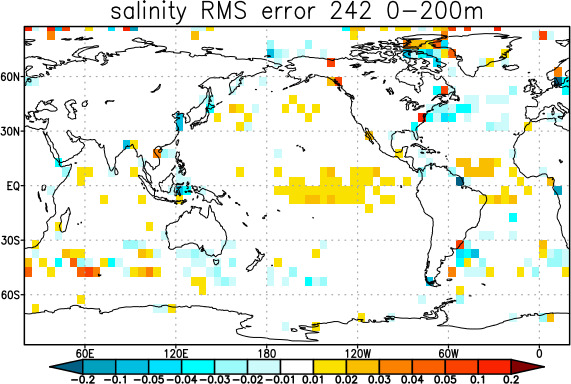}  &
            \includegraphics[width=0.35\linewidth]{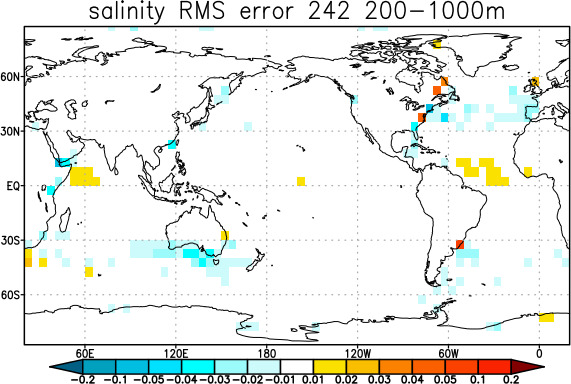} 
            \end{tabular}
            \caption{Top row: impact of using hybrid variances in $\mathbf{B}$  on the temperature RMS background errors in the top 200~m (left panel) and in the depth-range 200--1000~m (right panel); bottom row: the same but for salinity. The errors are plotted relative to those of the baseline experiment, which employed a parameterized $\mathbf{B}$.}
            \label{fig:rmse-map-hyb-var-ts}
        \end{figure}

\begin{figure}
            \centering
            \includegraphics[width=0.35\linewidth]{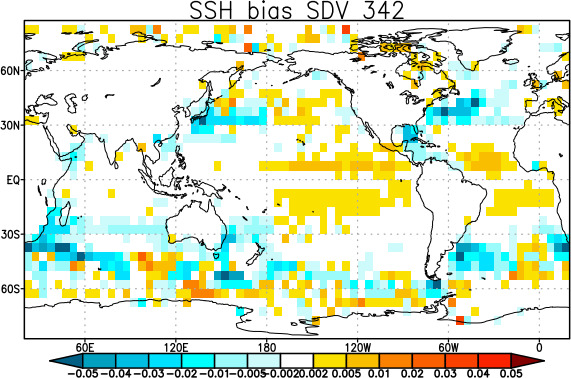}      \\
            \caption{Impact of using hybrid variances in $\mathbf{B}$ on the standard deviation of SSH background errors. The errors are plotted relative to those of the baseline experiment, which employed a parameterized $\mathbf{B}$.}
            \label{fig:std-map-hyb-var-ssh}
        \end{figure}

     We focus first on evaluating the impact of replacing the parameterized variances with the newly developed hybrid variances that we described in \Cref{subsec:bhybclim-var}. The diffusion tensor is the same parameterized tensor used in the baseline experiment. \Cref{fig:rmse-map-hyb-var-ts} shows spatial maps of the relative change in RMS background errors for temperature and salinity averaged over the two depth ranges of 0--200~m and 200--1000~m. For temperature, the results are largely positive in the extra-tropics, in particular in regions where the climatological parameters are able to describe more faithfully the ocean variability, specifically in the WBC and ACC regions. Some degradation is visible in the tropics, both in the Atlantic Ocean and Pacific Ocean. The hybrid variances, which by construction are larger than the parameterized values, allow the system to draw closely to observations in these regions. The results for salinity are more mixed. The RMS errors are reduced in the extra-tropics, but are also increased in the tropical Pacific Ocean and tropical Atlantic Ocean. There is a sign of degradation in the Antarctic region too. \Cref{fig:std-map-hyb-var-ssh} shows the relative change in the standard deviation of background errors for SSH. While the errors are significantly reduced in the WBC and ACC regions, there is also large degradation in the tropics and south of the ACC. As can be seen, using the hybrid variances in isolation brings a mixed bag of results. While the overall performance is positive, especially in the extra-tropics, noticeable degradation is also present. It should be noted that the hybrid variances were computed together with the hybrid tensor and are thus representative of the error structures induced by the latter. Still, we find it of interest to assess the impact of hybrid variances in isolation in order to better understand where the improvements associated with the hybrid $\mathbf{B}$ come from. In particular, such results may prove useful when trying to explain the origins of degradation, should there be any, of the full hybrid $\mathbf{B}$.

    \subsection{Impact of the hybrid tensor} 
    \label{subsec:bhybten-res}

    As described in \Cref{subsec:bhybclim-ten}, the hybrid correlation matrix is built from  a horizontal diffusion operator equipped with a climatological diffusion tensor and a vertical diffusion operator equipped with a fully flow-dependent vertical diffusion tensor. The vertical correlation length-scales diagnosed from the forecast ensemble are typically much larger in the mixed layer and in the deep ocean compared to those from the parameterized formulation. The larger values in the mixed layer should in principle allow the SST observations to have larger impact on the analysis by projecting them further into the mixed layer. \Cref{fig:rmse-map-hyb-ten-ts} shows spatial maps of the relative change in RMS {background errors for temperature and salinity for the two depth ranges of 0--200m and 200--1000m. For temperature in the extra-tropics, we can see a similar pattern to that obtained with the hybrid variances, where the errors are significantly reduced in the eddy-active regions. Interestingly, the background errors are now also reduced in the tropics, in particular in the upper ocean. Unlike when using hybrid variances, the background errors are larger in Baffin Bay at both plotted depth ranges. Salinity forecast errors are reduced in the tropics, with the exception of the Banda Sea, and the high-latitude extra-tropics. There is significant degradation in Baffin Bay and in the Arctic Ocean. Some degradation is also visible off the coasts of Antarctica. \Cref{fig:std-map-hyb-ten-ssh} shows the relative change in the standard deviation of background errors for SSH. The background errors are significantly reduced everywhere except in Baffin Bay and the Arctic Ocean. The degradation in all three variables at high latitudes may be associated with the relatively small horizontal climatological correlation length-scales compared to the parameterized length-scales, which is evident in \Cref{fig:parm-clim-ten-t} and \Cref{fig:parm-clim-ten-ssh}, for temperature and SSH, respectively. 

        \begin{figure}
            \centering
            \begin{tabular}{c c}
            \includegraphics[width=0.35\linewidth]{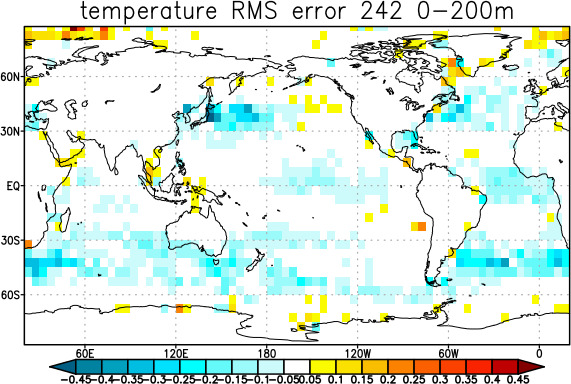}  &
            \includegraphics[width=0.35\linewidth]{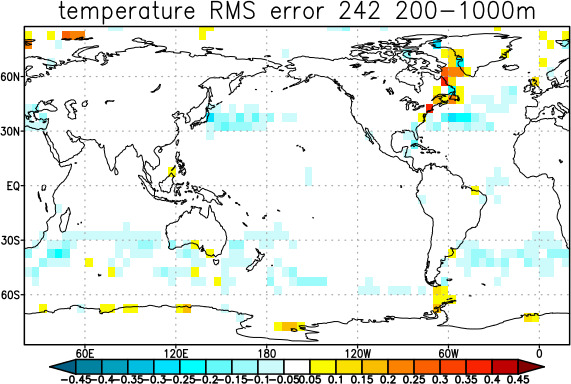}            \vspace{0.25cm} \\
            \includegraphics[width=0.35\linewidth]{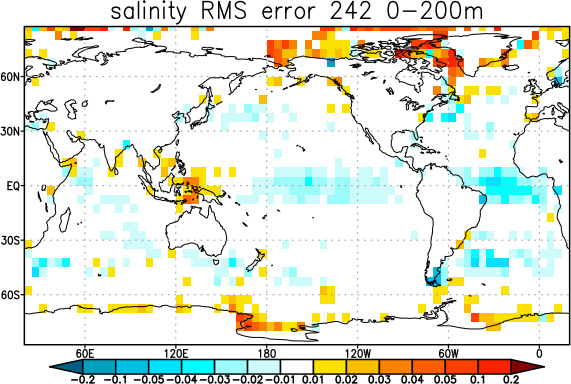}  &
            \includegraphics[width=0.35\linewidth]{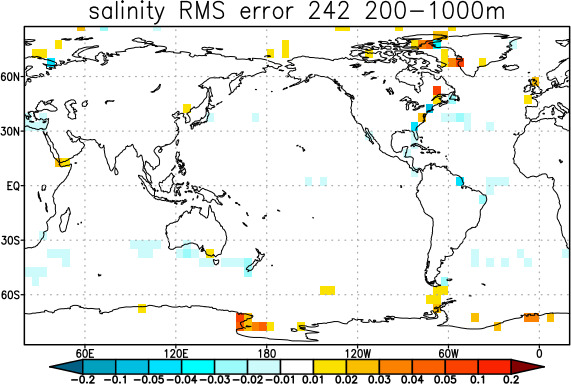} 
            \end{tabular}
            \caption{Top row: impact of using the hybrid diffusion tensor in $\mathbf{B}$ on the temperature RMS background errors in the top 200~m (left panel) and in the depth-range 200--1000m (right panel); bottom row: the same but for salinity. The errors are plotted relative to those of the baseline experiment, which employed a parameterized $\mathbf{B}$.}
            \label{fig:rmse-map-hyb-ten-ts}
        \end{figure}

        \begin{figure}
            \centering
            \includegraphics[width=0.35\linewidth]{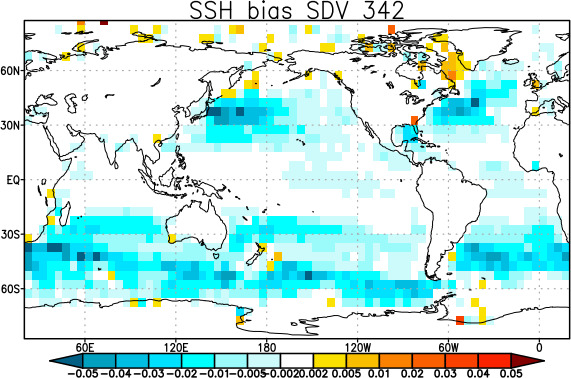}      \\
            \caption{Impact of using the hybrid diffusion tensor in $\mathbf{B}$ on the standard deviation of SSH background errors. 
            The errors are plotted relative to those of the baseline experiment, which employed a parameterized $\mathbf{B}$.}
            \label{fig:std-map-hyb-ten-ssh}
        \end{figure}

    \subsection{Impact of the hybrid variances and hybrid tensor combined}
    \label{subsec:bhybtenvar-res}

        \begin{figure}
            \centering
            \begin{tabular}{c}
            \includegraphics[width=0.65\linewidth]{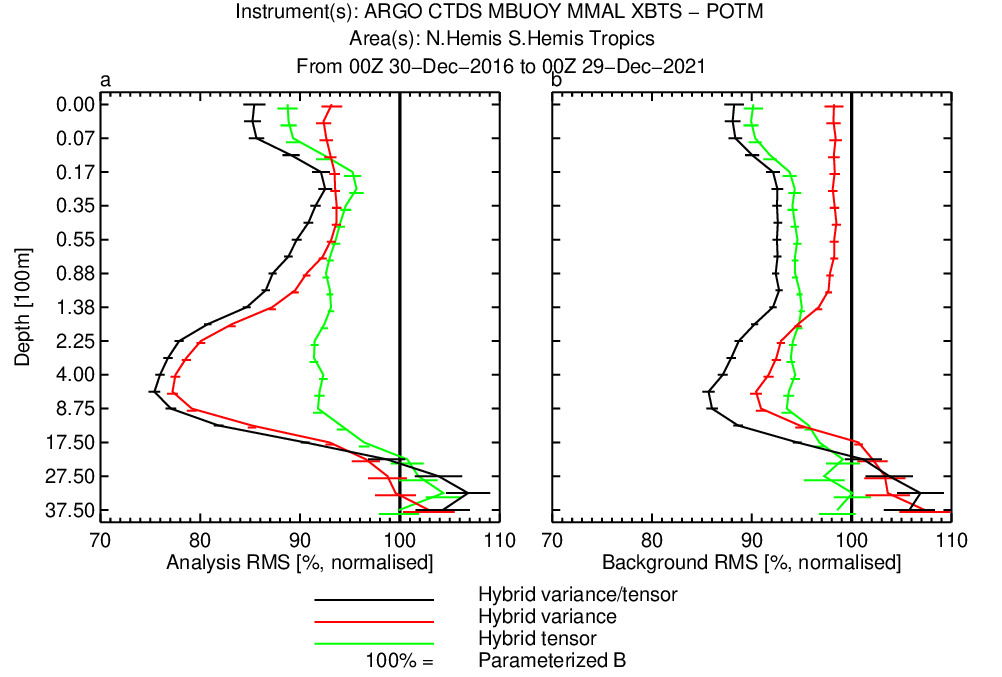}      \\
            \end{tabular}
            \caption{Globally averaged normalized RMS errors for the analysis (left panel) and background (right panel) computed with respect to the assimilated observations for temperature as a function of ocean depth. The baseline experiment with the parameterized formulation of $\mathbf{B}$  marks the 100\% line. }
            \label{fig:rmse-prof-hyb-varten-t}
        \end{figure}

        \begin{figure}
            \centering
            \begin{tabular}{c}
            \includegraphics[width=0.65\linewidth]{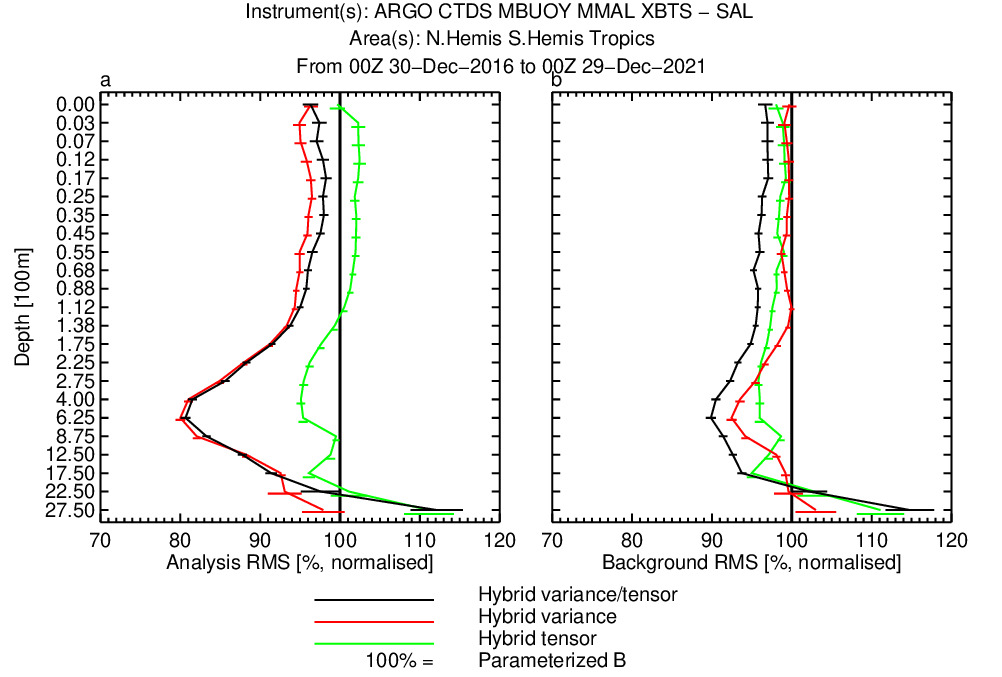}      \\
            \end{tabular}
            \caption{Globally averaged normalized RMS errors for the analysis (left panel) and background (right panel) computed with respect to the assimilated observations for salinity as a function of ocean depth. The baseline experiment with the parameterized $\mathbf{B}$ marks the 100\% line.}
            \label{fig:rmse-prof-hyb-varten-s}
        \end{figure}

        \begin{figure}
            \centering
            \begin{tabular}{c}
            \includegraphics[width=0.65\linewidth]{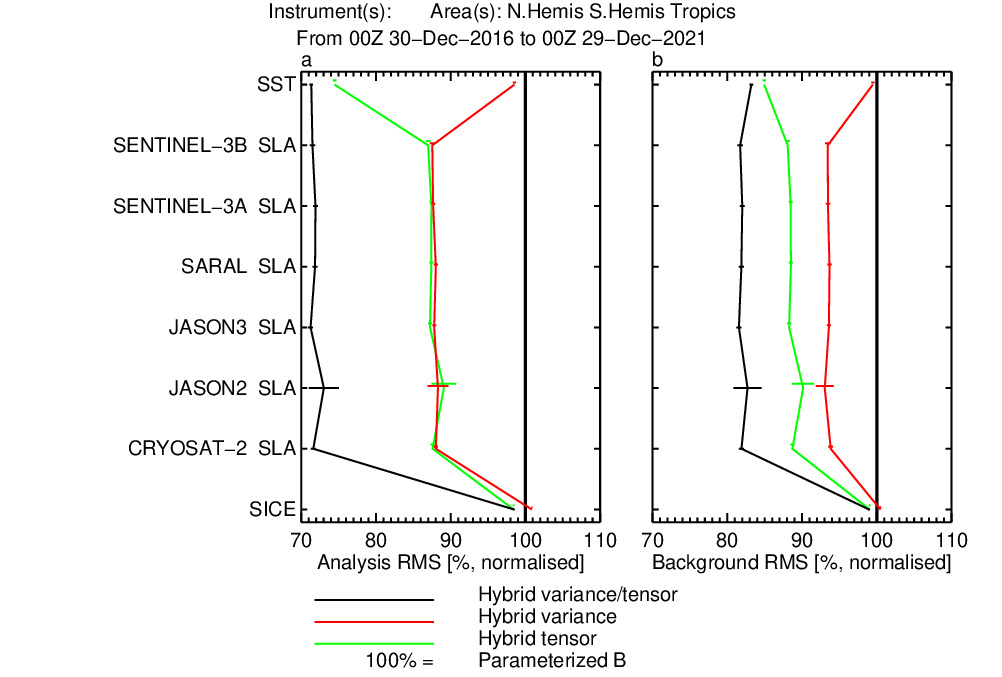}      \\
            \end{tabular}
            \caption{Globally averaged normalized RMS errors for the analysis (left panel) and background (right panel) computed with respect to the assimilated observations for SST and SLA. The baseline experiment with the parameterized formulation of $\mathbf{B}$ marks the 100\% line.}
            \label{fig:rmse-hyb-varten-ssh}
        \end{figure}

    We demonstrated in the previous subsections that the impact of the hybrid variances and the hybrid tensor on reducing background  errors is generally positive when each of them is tested independently from the other. There were some caveats, however. In both cases, the impact is largely positive in the WBC and ACC regions, which are important regions from the point of view of medium-range coupled ocean-atmosphere forecasting. We now check the performance of using the hybrid variances together with the hybrid tensor. 
    
    In order to provide a more quantitative assessment of the impact, we show in \Cref{fig:rmse-prof-hyb-varten-t} the globally averaged vertical profiles of the normalized RMS errors of the analysis and background computed against the assimilated observations for temperature for the three considered configurations: with hybrid variances, hybrid diffusion tensor and both together. The RMS errors are normalized with the RMS errors of the baseline experiment employing the parameterized $\mathbf{B}$.  As could be anticipated, applying hybrid variances while keeping the parameterized tensor results in an analysis that fits much more closely the observations. This is consistent with the fact that the climatological variances are larger in the eddy-active regions around the globe. The improvement of the background fit to observations, on the other hand, is relatively modest, especially in the upper ocean. This may be due either to over-fitting the observations or mis-specifying the correlation length-scales. When applying the hybrid tensor together with parameterized variances, on the other hand, the background fit to the observations is improved to a similar extent as the analysis fit to the observations. It may be expected that both drawing more to observations in the analysis and improving the representation of the structure of the correlation matrix should be beneficial. This is indeed the case as the configuration with both hybrid variances and hybrid tensor performs the best. The globally averaged temperature RMS errors are reduced by roughly $10\%$ throughout the whole water column with respect to the configuration employing the ORAS5 $\mathbf{B}$. 
    
    \Cref{fig:rmse-prof-hyb-varten-s} shows the corresponding globally averaged vertical profiles of normalized RMS errors for salinity. This time, using the hybrid tensor alone results in a degraded analysis fit to observations in the upper 1000~m, while the background fit is only marginally improved. Employing hybrid variances results in the analysis drawing closer to observations, but with little impact on the background fit to observations. Only when the hybrid variances are combined with the hybrid tensor, can we see a significant improvement to the system's performance. The background fit to salinity observations is improved by over $5\%$ in the upper ocean and up to $10\%$ around a depth of 600~m. \Cref{fig:rmse-hyb-varten-ssh} compares the globally averaged analysis and background fit to SST and SLA observations for the three configurations. Once again, while both the hybrid variance and hybrid tensor configurations show reduced analysis and background errors, it is when combining both hybrid elements that we see the best performance. Both the background fit to SST and SLA observations is improved by nearly $20\%$.

        \begin{figure}
            \centering
            \begin{tabular}{c c}
            \includegraphics[width=0.35\linewidth]{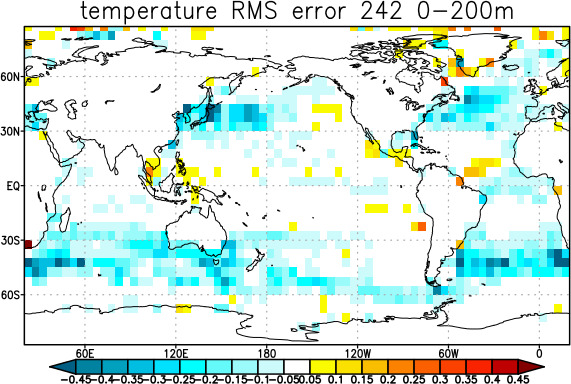}  &
            \includegraphics[width=0.35\linewidth]{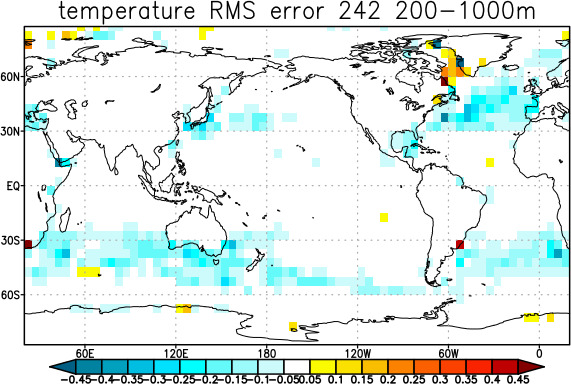}            \vspace{0.25cm} \\
            \includegraphics[width=0.35\linewidth]{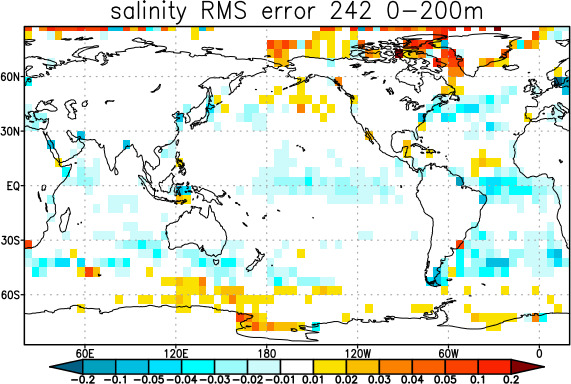}  &
            \includegraphics[width=0.35\linewidth]{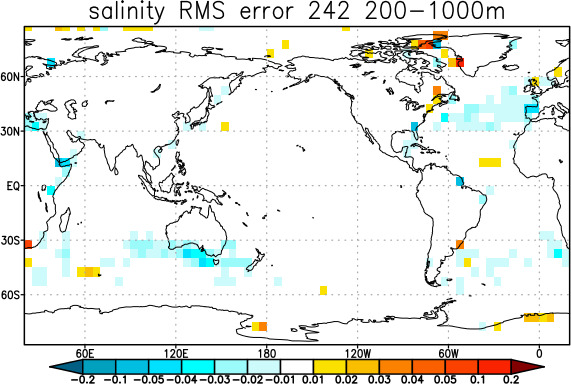} 
            \end{tabular}
            \caption{Top row: impact of using the hybrid $\mathbf{B}$ on the temperature RMS background errors in the top 200~m (left panel) and in the depth-range 200--1000~m (right panel); bottom row: the same but for salinity.}
            \label{fig:rmse-map-hyb-varten-ts}
        \end{figure}
        
        \begin{figure}
            \centering
            \includegraphics[width=0.35\linewidth]{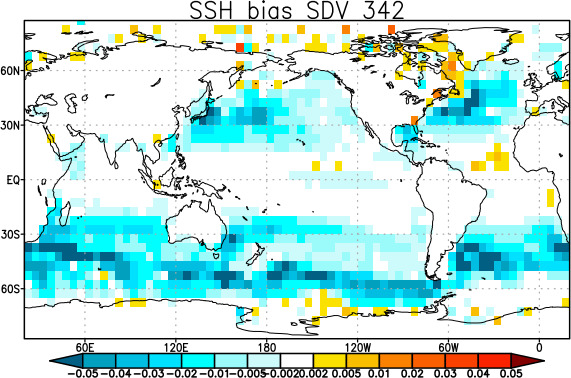}      \\
            \caption{Impact of using the hybrid tensor in $\mathbf{B}$ on the standard deviation of the background errors for SSH.}
            \label{fig:std-map-hyb-varten-ssh}
        \end{figure}
       
    \Cref{fig:rmse-map-hyb-varten-ts} shows spatial maps of the relative RMS background errors for temperature and salinity for the two depth ranges of 0--200m and 200--1000m. The temperature RMS errors are significantly reduced in the WBC and ACC regions. The degradation in Baffin Bay, also observed when using the hybrid tensor in isolation, is also visible. The salinity background errors are significantly reduced in the upper ocean apart from high-latitude regions in the Northern and Southern Hemispheres, which, as remarked earlier, may be associated with shorter horizontal correlation length-scales in the hybrid tensor. \Cref{fig:std-map-hyb-varten-ssh} shows the relative change in the standard deviation of background errors for SSH. Apart from impressive background-error reduction in eddy-active regions, Baffin Bay stands out as an area where the background errors are increased. 

    The location of large reductions in the background errors correlates well with areas where the hybrid variances are larger compared to their parameterized counterparts. As could be appreciated from figures showing vertical profiles of RMS analysis and background errors, the hybrid variances only explain part of the improvements. It is the hybrid tensor, with better representation of correlation spatial structure, that brings a significant impact. While the horizontal correlation length-scales are better aligned with the boundaries of ocean currents, the flow-dependent specification of vertical correlation length-scales plays an even more significant role, in particular in improving the vertical projection of information coming from SST observations. The increments obtained when using the parameterized formulation of $\mathbf{B}$ are relatively shallow and therefore incapable of correcting background errors throughout the mixed layer. The flow-dependent specification of vertical correlation length-scales, on the other hand, results in increments extending to the bottom of the mixed layer and thus provides a more effective correction to the upper ocean.


    The inability of the parameterized $\mathbf{B}$ to project the SST increments into the mixed layer causes large model biases to develop in a cycled data assimilation system. This problem is particularly acute in the Gulf Stream area. Since IFS cycle 45r1, ECMWF's medium-range forecasts are fully coupled to OCEAN5 in the tropics, but are only partially coupled in the extra-tropics \citep{browne-2019}. This partial coupling strategy is used because of a known deficiency of ORAS5: it is not able to accurately represent the position of western boundary currents. \Cref{fig:sst-bias} shows the SST bias computed for the period 2017--2021 when using the parameterized $\mathbf{B}$ (left panel) and when using the new hybrid $\mathbf{B}$. The bias is largely reduced, thus mitigating the need for a partial coupling strategy in ORAS6.

        \begin{figure}
           \centering
           \begin{tabular}{c c}
            \includegraphics[width=0.45\linewidth]{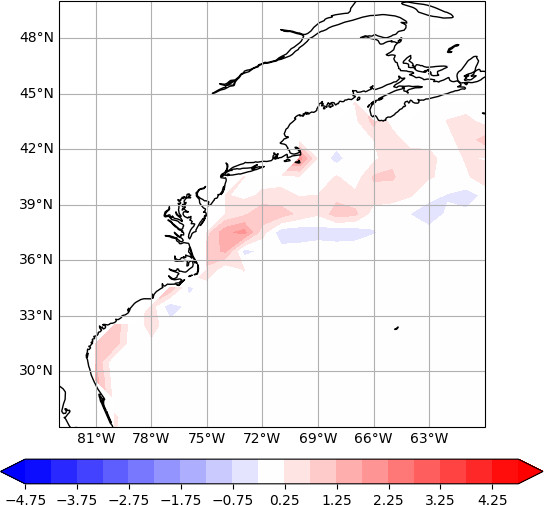} & \includegraphics[width=0.45\linewidth]{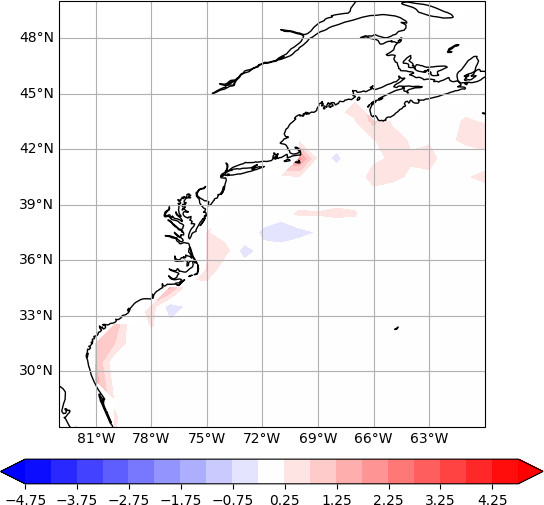} 
            \end{tabular}
            \caption{SST bias (in K) obtained when using the parameterized $\mathbf{B}$ (left panel) and the new hybrid $\mathbf{B}$ (right panel). SST is verified against the European Space Agency CCI2 SST dataset in the period 2017--2021.}
            \label{fig:sst-bias}
        \end{figure}
        
\section{Conclusions} \label{sec:conclusions}

In this article, we have described an Ensemble of Data Assimilations framework that has been developed for ECMWF's next generation ocean reanalysis system, ORAS6. The forecast ensemble is used in two ways: first, to construct seasonal climatologies of parameters of the background-error covariance matrix (${\bf B}$), specifically the variances and the horizontal and vertical length-scales of the correlation model; and second, to provide flow-dependent estimates, on each assimilation cycle, of the background-error variances and vertical correlation length-scales. The climatological and flow-dependent estimates are blended in a hybrid formulation of $\mathbf{B}$. The impact of this new, hybrid $\mathbf{B}$ has been assessed with respect to the formulation of $\mathbf{B}$ used in OCEAN5 and ORAS5, ECMWF's current ocean analysis and reanalysis systems. In the current system, the variances and correlation length-scales are  parameterized using empirical and analytical relationships, without making use of an ensemble. 

The performance of the new formulation of $\mathbf{B}$ has been assessed in the Argo-rich period. The climatological estimates of the covariance parameters have also been computed in the Argo-rich period, over a 6-year interval preceding the assessment period. The new formulation of $\mathbf{B}$ has not been assessed, however, in periods prior to the introduction of Argo. It is worth remarking that the climatological estimates of the covariance parameters computed here are not appropriate for the pre-Argo period when background errors can be expected to be much larger. Applying the hybrid $\mathbf{B}$ in that period, or in any period where there is a substantial change in the observation network, should require carefully revising the climatology.

The results from this study are overwhelmingly positive with the new, hybrid formulation of $\mathbf{B}$, with background errors being significantly reduced for temperature, salinity and sea surface height.  Improvements are largest in extra-tropical eddy-active regions. Relatively small degradation is present at high latitudes, in particular in Baffin Bay and in the Arctic Ocean. An important feature of the new $\mathbf{B}$ is that it allows increments from SST observations to be projected effectively into the mixed layer, which is not possible with the parameterized $\mathbf{B}$ used in ORAS5. As a result, in ORAS6, SST observations can now be assimilated directly using the 3D-Var FGAT scheme, and hence in synergy with the other assimilated observations, instead of indirectly via an {\it ad hoc} nudging scheme as done in ORAS5.

There is considerable scope for improving the EDA, both in the ensemble-generation strategy and in the way the ensembles are used in the hybrid $\mathbf{B}$. Improving the quality and reliability of the forecast ensemble can potentially be achieved by introducing stochastic perturbations in the ocean model to account for  uncertainty arising from random model errors. The covariance model can also be enhanced to make more effective use of the ensemble information. For example, the diffusion-based correlation model could be generalized to account for a non-diagonal diffusion tensor in order to exploit the anisotropic information in the ensemble more appropriately than done here. Building scale dependence into the covariance model, so that observations can correct different spatial scales more effectively, is another promising avenue to explore. Recent work to develop a scale-dependent, ensemble-based $\mathbf{B}$ suggests that this approach can be substantially beneficial to the analysis. 



\section*{Acknowledgements}

This work has benefited from funding from the Copernicus Climate Change Service (C3S), one of six services of the European Union’s Copernicus Programme, which is implemented by ECMWF on behalf of the European Commission. In particular, this work is a contribution to  contract C3S\_321b\_INRIA: ``Technical preparations for C3S Seasonal Initialization and Global Reanalysis: Enabling an Ensemble of Data Assimilations for the Ocean''. We would like to thank Massimo Bonavita and Tony McNally for their comments and suggestions to improve the manuscript.

\bibliography{bibtex}

\end{document}